\documentclass[3p,review]{elsarticle}

\usepackage[english]{babel}
\usepackage{units}
\usepackage{graphicx}
\usepackage{amsmath}
\usepackage{amssymb}
\usepackage{subfigure}
\usepackage{lineno}

\begin{document}

\begin{frontmatter}

\title{Measuring a Cherenkov ring in the radio emission from air showers at 110-190 MHz with LOFAR}

\author[ru,ni]{A.~Nelles}
\cortext[mycorrespondingauthor]{Corresponding author}
\ead{a.nelles@astro.ru.nl}
\author[ru]{P.~Schellart}
\author[ru]{S.~Buitink}
\author[ru]{A.~Corstanje}
\author[brussels]{K.~D.~de Vries}
\author[ru]{J.~E.~Enriquez}
\author[ru,as,ni,mpifr]{H.~Falcke}
\author[as]{W.~Frieswijk}
\author[ru,ni]{J.~R.~H\"orandel}
\author[rug]{O.~Scholten}
\author[ru]{S.~ter Veen}
\author[ru]{S.~Thoudam}
\author[ru]{M.~van den Akker}
\author[aip]{J.~Anderson}
\author[astron,shell]{A.~Asgekar}
\author[caastro]{M.~E.~Bell}
\author[astron]{M.~J.~Bentum}
\author[cfa]{G.~Bernardi}
\author[roe]{P.~Best}
\author[astron]{J.~Bregman}
\author[aip]{F.~Breitling}
\author[soton]{J.~Broderick}
\author[astron,kapteyn]{W.~N.~Brouw}
\author[hamburg]{M.~Br\"uggen}
\author[anu]{H.~R.~Butcher}
\author[mpifa]{B.~Ciardi}
\author[astron]{A.~Deller}
\author[astron]{S.~Duscha}
\author[tls]{J.~Eisl\"offel}
\author[astron]{R.~A.~Fallows}
\author[astron,leiden]{M.~A.~Garrett}
\author[astron]{A.~W.~Gunst}
\author[soton,jod]{T.~E.~Hassall}
\author[astron]{G.~Heald}
\author[mpifr]{A.~Horneffer}
\author[leiden]{M.~Iacobelli}
\author[raiub]{E.~Juette}
\author[ox]{A.~Karastergiou}
\author[astron,lebedev]{V.~I.~Kondratiev}
\author[mpifr,jod]{M.~Kramer}
\author[mpifr]{M.~Kuniyoshi}
\author[astron]{G.~Kuper}
\author[astron]{P.~Maat}
\author[aip]{G.~Mann}
\author[astron,kapteyn]{M.~Mevius}
\author[astron]{M.~J.~Norden}
\author[groningen]{H.~Paas}
\author[lyon]{M.~Pandey-Pommier}
\author[ox]{G.~Pietka}
\author[astron]{R.~Pizzo}
\author[astron]{A.~G.~Polatidis}
\author[mpifr]{W.~Reich}
\author[leiden]{H.~R\"ottgering}
\author[soton]{A.~M.~M.~Scaife}
\author[bielefeld]{D.~Schwarz}
\author[crat,skasa]{O.~Smirnov}
\author[jod]{B.~W.~Stappers}
\author[aip]{M.~Steinmetz}
\author[ox]{A.~Stewart}
\author[cnrs]{M.~Tagger}
\author[astron]{Y.~Tang}
\author[meudon]{C.~Tasse}
\author[astron]{R.~Vermeulen}
\author[aip]{C. Vocks}
\author[cfa]{R.~J.~van Weeren}
\author[astron]{S.~J.~Wijnholds}
\author[ubonn,mpifr]{O.~Wucknitz}
\author[astron]{S.~Yatawatta}
\author[meudon]{P.~Zarka}

\address[ru]{Department of Astrophysics/IMAPP, Radboud University Nijmegen, P.O. Box 9010, 6500 GL Nijmegen, The Netherlands}
\address[ni]{Nikhef, Science Park Amsterdam, 1098 XG Amsterdam, The Netherlands}
\address[brussels]{Vrije Universiteit Brussel, Dienst ELEM, B-1050 Brussels, Belgium}
\address[as]{Netherlands Institute for Radio Astronomy (ASTRON), Postbus 2, 7990 AA Dwingeloo, The Netherlands}
\address[mpifr]{Max-Planck-Institut f\"{u}r Radioastronomie, Auf dem H\"ugel 69, 53121 Bonn, Germany}
\address[rug]{KVI, University Groningen, 9747 AA Groningen, The Netherlands}
\address[aip]{Leibniz-Institut f\"{u}r Astrophysik Potsdam (AIP), An der Sternwarte 16, 14482 Potsdam, Germany }
\address[astron]{Netherlands Institute for Radio Astronomy (ASTRON), Postbus 2, 7990 AA Dwingeloo, The Netherlands }
\address[shell]{Shell Technology Center, Bangalore, India }
\address[caastro]{ARC Centre of Excellence for All-sky astrophysics (CAASTRO), Sydney Institute of Astronomy, University of Sydney Australia }
\address[cfa]{Harvard-Smithsonian Center for Astrophysics, 60 Garden Street, Cambridge, MA 02138, USA }
\address[roe]{Institute for Astronomy, University of Edinburgh, Royal Observatory of Edinburgh, Blackford Hill, Edinburgh EH9 3HJ, UK }
\address[soton]{School of Physics and Astronomy, University of Southampton, Southampton, SO17 1BJ, UK }
\address[kapteyn]{Kapteyn Astronomical Institute, PO Box 800, 9700 AV Groningen, The Netherlands }
\address[hamburg]{University of Hamburg, Gojenbergsweg 112, 21029 Hamburg, Germany }
\address[anu]{Research School of Astronomy and Astrophysics, Australian National University, Mt Stromlo Obs., via Cotter Road, Weston, A.C.T. 2611, Australia }
\address[mpifa]{Max Planck Institute for Astrophysics, Karl Schwarzschild Str. 1, 85741 Garching, Germany }
\address[tls]{Th\"{u}ringer Landessternwarte, Sternwarte 5, D-07778 Tautenburg, Germany }
\address[leiden]{Leiden Observatory, Leiden University, PO Box 9513, 2300 RA Leiden, The Netherlands }
\address[jod]{Jodrell Bank Center for Astrophysics, School of Physics and Astronomy, The University of Manchester, Manchester M13 9PL,UK }
\address[raiub]{Astronomisches Institut der Ruhr-Universit\"{a}t Bochum, Universit\"{a}tsstra{\ss}e 150, 44780 Bochum, Germany }
\address[ox]{Astrophysics, University of Oxford, Denys Wilkinson Building, Keble Road, Oxford OX1 3RH }
\address[lebedev]{Astro Space Center of the Lebedev Physical Institute, Profsoyuznaya str. 84/32, Moscow 117997, Russia }
\address[groningen]{Center for Information Technology (CIT), University of Groningen, The Netherlands }
\address[lyon]{Centre de Recherche Astrophysique de Lyon, Observatoire de Lyon, 9 av Charles Andr\'{e}, 69561 Saint Genis Laval Cedex, France }
\address[bielefeld]{Fakult\"{a}t f\"{u}r Physik, Universit\"{a}t Bielefeld, Postfach 100131, D-33501, Bielefeld, Germany }
\address[crat]{Centre for Radio Astronomy Techniques \& Technologies (RATT), Department of Physics and Elelctronics, Rhodes University, PO Box 94, Grahamstown 6140, South Africa }
\address[skasa]{SKA South Africa, 3rd Floor, The Park, Park Road, Pinelands, 7405, South Africa }
\address[cnrs]{LPC2E - Universite d'Orleans/CNRS }
\address[meudon]{LESIA, UMR CNRS 8109, Observatoire de Paris, 92195   Meudon, France }
\address[ubonn]{Argelander-Institut f\"{u}r Astronomie, University of Bonn, Auf dem H\"{u}gel 71, 53121, Bonn, Germany }

\begin{abstract}
Measuring radio emission from air showers offers a novel way to determine properties of the primary cosmic rays such as their mass and energy. Theory predicts that relativistic time compression effects lead to a ring of amplified emission which starts to dominate the emission pattern for frequencies above $\unit[\sim100]{MHz}$. In this article we present the first detailed measurements of this structure. Ring structures in the radio emission of air showers are measured with the LOFAR radio telescope in the frequency range of $\unit[110-190]{MHz}$. These data are well described by CoREAS simulations. They clearly confirm the importance of including the index of refraction of air as a function of height. Furthermore, the presence of the Cherenkov ring offers the possibility for a geometrical measurement of the depth of shower maximum, which in turn depends on the mass of the primary particle.
\end{abstract}

\end{frontmatter}

\section{Introduction}
Since the discovery of radio emission from cosmic ray induced air showers in the 1960s \cite{Jelley1966,Allan:1966} it has been predicted that radio emission can be used as a tracer for arrival direction, mass, and energy of the primary particle. Experimentally it has been shown that the amplitude of the radio pulse scales with the energy of the primary particle \cite{Falcke:2005}. Recently, it has also been demonstrated that the radio emission pattern observed at the ground is sensitive to the atmospheric depth of shower maximum $X_\mathrm{max}$ \cite{Buitink2014,LopesMuonHeight2013}. This parameter which depends on the type or mass of the primary particle can be measured with an accuracy comparable to that of established techniques, such as the detection of Fluorescence or Cherenkov light (e.g.\cite{KampertUnger}), but with a much better duty cycle. 

This progress would not have been made without thorough theoretical understanding of the emission processes. Recently, models describing the emission processes have converged to similar results \cite{ComparisonREASMGMR}. The primary emission component is caused by deflection of electrons and positrons in the Earth's magnetic field, producing a coherent radio pulse that is polarized linearly in a direction perpendicular to the magnetic field direction and the direction of propagation of the air shower. A secondary component due to negative charge excess at the shower front also generates a coherent radio pulse that is polarized linearly, but now radially away from the shower axis. These components either add constructively or destructively depending on the observer location, creating a complex emission pattern at ground level \cite{CoREAS}. This pattern is further influenced by the non-constant and non-unity refractive index of the atmosphere, which leads to visible effects of relativistic time compression of the signal. Hereby emission is amplified for a specific angle with respect to the shower axis \cite{Vries:2012,Alvarez-Muniz:2012}. Time compression of the pulse also allows shower emission to be detectable at GHz frequencies where coherence would otherwise be lost \cite{Anita2010,Smida:2013}.

The predicted observational signature of the relativistic time compression of the signal is a ring of amplified emission at the Cherenkov angle, with a diameter in the order of $\unit[100]{m}$ depending on the shower geometry \cite{Vries:2012}. For frequencies  above $\unit[100]{MHz}$ this ring is predicted to dominate the emission pattern, but this has yet to be observationally confirmed. From geometrical considerations it follows that the ring diameter should trace the distance to the shower maximum. This can be combined with atmospheric models to derive $X_{\mathrm{max}}$. Simulation studies confirm this basic idea \cite{de-Vries:2013}. Thus, measuring the ring diameter might provide an alternate, geometrical method to derive $X_{\mathrm{max}}$, which can be combined with air shower models to reconstruct the type of primary particle.

In this article, measurements of radio emission from cosmic ray air showers in the frequency range $\unit[110-190]{MHz}$ are presented. While this frequency range has been probed inconclusively by several early experiments \cite{Jelley1966,Charman1969,Spencer1969} and other recent experiments are also sensitive in this band \cite{Codalema2009}, these data represent the first high-quality measurements of the radio pattern on a single event basis. We present three example events that show for the first time on a single shower basis the importance of the Cherenkov ring in this frequency range. Subsequently we test the possibility of using the radius of the Cherenkov ring as an estimator for the depth of shower maximum.

The instrument is described in section~\ref{sec:instrument}, followed by the description of employed data analysis techniques in section~\ref{sec:detection}. The current dataset is characterized in section~\ref{sec:dataset}. Finally, in section~\ref{sec:discussion} densely sampled signal patterns are shown and their implications for $X_\mathrm{{max}}$ measurements are discussed.

\section{The instrument}
\label{sec:instrument}
\begin{figure}
\includegraphics[width=0.48\textwidth]{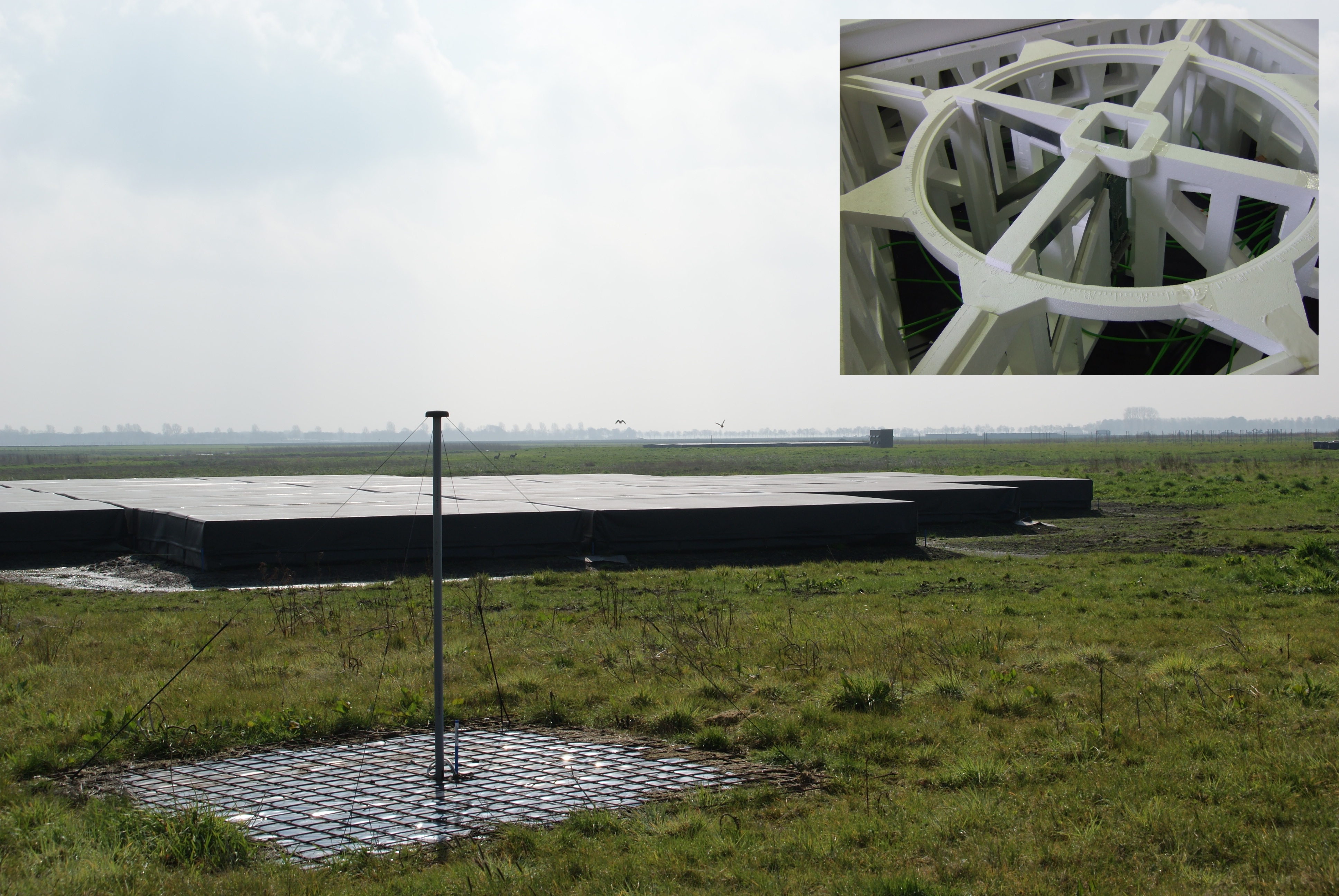}
	\caption[HBA Design]{Radio antennas at LOFAR. Behind a low-band antenna a cluster of 24 black tiles of high-band antennas are shown. The inset shows the construction of a high-band element in which the bow-tie shaped antennas are mounted before they are packed in weather-proof foil. }
	\label{fig:hba-antenna}
\end{figure}

LOFAR is a distributed radio telescope targeted at observing the lowest radio frequencies from $\unit[240]{MHz}$ down to the atmospheric cut-off at $\unit[10]{MHz}$. The antennas of LOFAR are distributed over several European countries with a dense core in the Netherlands. The instrument was specifically designed to be able to observe short duration radio signals, such as those emitted by pulsars or cosmic ray induced air showers \cite{LOFAR}.

LOFAR covers the low frequency range with two types of antennas. The low-band antennas (\emph{LBAs}, $\unit[10-90]{MHz}$) consist of two inverted V-shaped dipole antennas, which are read-out individually. Their characteristics with respect to cosmic-ray air shower measurements are described in \cite{Schellart2013}.

The high-band antennas (\emph{HBAs}) cover the frequency range from $\unit[110-240]{MHz}$. In this range the antennas are no longer sky-noise dominated, which had to be accounted for in the design, while keeping the antennas as low cost as possible. One HBA \emph{element} consist of dual-polarization fat dipole \emph{antennas}, in which holes were cut to save material, making them similar to bow-tie antennas. The elements are arranged in a styrofoam structure and combined in groups of 16, called \emph{tiles}. Every tile is packed in black foil, to protect the antenna electronics from the weather. Examples of an element and a tile can be seen in figure \ref{fig:hba-antenna}.

The HBA electronics have been optimized for targeted astronomical observations. The signals from all antennas in a tile are amplified and combined in an analog beamforming step. This means that the signals from the antennas are no longer available separately, but summed with a correction for a delay that a source from a certain direction would introduce. The delays to be applied are provided by the LOFAR control system for a user selectable direction. They are updated very $\unit[180]{s}$ to keep the direction of maximum sensitivity pointing in the same direction during an observation. A maximum delay of $\unit[15.5]{ns}$ can be introduced in $32$ steps \cite{Kant2007}.

\begin{figure}
\includegraphics[width=0.48\textwidth]{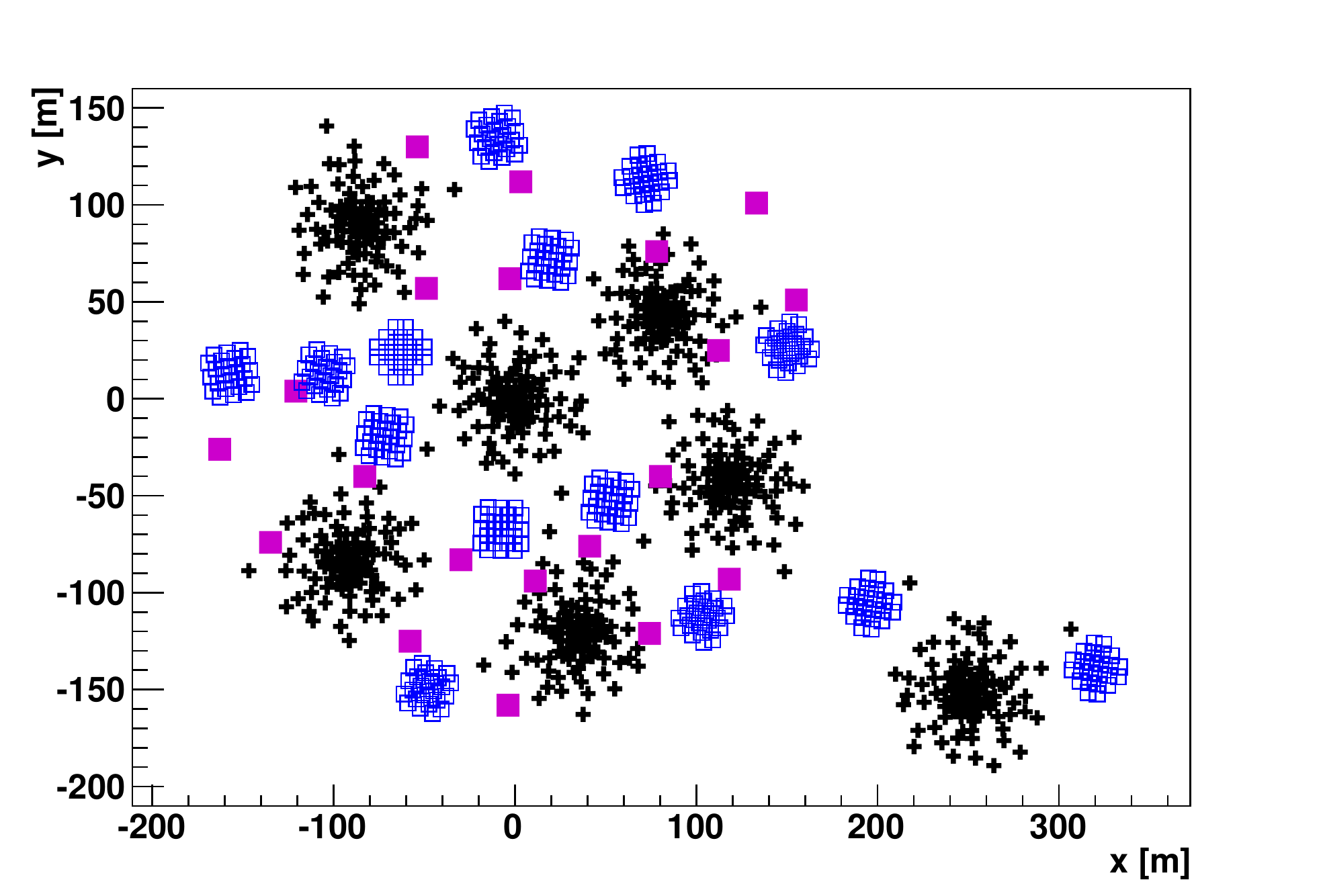}
	\caption[Superterp]{Central core of LOFAR. The black crosses are the low-band antennas, the blue open squares indicate the tiles of high-band antennas and the filled squares are the particle detectors.}
	\label{fig:superterp}
\end{figure}

The HBAs (and LBAs) are grouped together in \emph{stations} which themselves are distributed on an irregular grid to maximize the number of different baselines for interferometric observations. Near the center of LOFAR two groups of 24 tiles, called \emph{sub-stations}, are associated with every station as indicated in figure \ref{fig:superterp}. Further away from the core, 48 tiles in a single group belong to a station and international stations comprise a group of 96 tiles. 

LOFAR is continuously performing astronomical observations. Here the primary observer selects a target to observe and the desired frequency band. Using both the low-band and high-band antennas simultaneously is currently not possible. Cosmic ray observations typically run in parallel to these primary observations, thus the amount of observing time in a given frequency band is not controlled by the cosmic ray project.

In parallel to any observation the tile-beamformed data from all tiles are filled into ring buffers (Transient Buffer Boards), from which the last $\unit[5]{s}$ of data can be recorded when triggered. Triggers can be generated by inspecting the data with an on-board FPGA\footnote{Field programmable gate array.} or received through the LOFAR control software. In order to record cosmic-ray pulses, the dense core of LOFAR is equipped with an array of particle detectors \cite{Thoudam2013}, as also shown in figure \ref{fig:superterp}. In routine observations, coincidences of several particle detectors trigger a read-out of the ring buffers \cite{Schellart2013}.

The HBAs can be sampled at two different clock frequencies, which allows for several different observing bands. The one mostly used in the present data-set is $\unit[110-190]{MHz}$ with $\unit[200]{MHz}$ sampling, i.e. second Nyquist zone. Alternatively, observations of $\unit[170-230]{MHz}$ ($\unit[160]{MHz}$, third Nyquist zone) or $\unit[210-240]{MHz}$ ($\unit[200]{MHz}$, third Nyquist zone) can be chosen. However, so far no cosmic-ray observations were conducted in the highest band.

% -----------------------------------
% -----------------------------------

\section{Detecting signals from cosmic rays}
\label{sec:detection}
The detection of cosmic rays in the data from the HBAs is performed mostly identical to that of the LBAs as described in detail in \cite{Schellart2013}. Here a brief overview is given, and the differences for HBA measurements are explained. Cosmic rays are detected in parallel to an ongoing astronomical observation, which determines the direction of analogue beamforming. When a trigger from the array of particle detectors is received, the tile-beamformed data is written to disk and stored for processing. The combination of all radio and particle data corresponding to one trigger is called an \emph{event}. During processing, the signals from all tiles in a station are first coherently beamformed in the cosmic-ray arrival direction as reconstructed from the particle data. When a significant signal, with a signal-to-noise ratio exceeding three in amplitude, is detected in the beamformed signal the station is selected for further processing and the event is selected as a cosmic ray candidate. Using a smaller search window, around the peak in the beamformed signal, a pulse search is then performed on the Hilbert envelope of the up-sampled signals from each tile. Up-sampling, by a factor $16$, is needed such that the pulse maximum search is not the limiting factor in achieving the required time resolution. From the arrival times of those pulse maxima for which the signal-to-noise ratio in amplitude exceeds three, the direction of the cosmic ray is reconstructed. Additionally the amplitude (in each instrumental polarization) and integrated pulse power, over $\unit[55]{ns}$ centered on the pulse maximum, are extracted.

However, due to the different hardware and other dominant contributions to the background there are some differences in the reconstruction of the HBA data. 
\subsection{Removal of radio frequency interference (RFI)}
The frequency band of the HBAs is less radio quiet than the LBA band. Especially an emergency pager signal at around 170 MHz adds a non-negligible amount of power to the spectrum. Therefore, before additional RFI removal is applied \cite{Schellart2013}, all power in a band of $\unit[3]{MHz}$ around the pager frequency is set to zero with the edges convolved with a Gaussian to prevent artificial ringing in the signal.

Furthermore, single HBAs have been reported to occasionally show spikes in the data due to malfunctions. As such spikes will disturb the initial search for pulses in a beam formed trace, a simple spike search is performed after the RFI cleaning and antennas with spikes of outlying high amplitude are removed from the data-set.
\subsection{Non-removable background}
\label{sec:back}
The HBAs are no longer fully dominated by the diffuse sky noise. In addition to the system noise, some astronomical sources introduce measurable signals in every single tile, most evident for bright sources such as Cas A or Cyg A. This means that the background noise in HBA observation is neither uniform nor independent of the direction of observation.

With dedicated on- and off-source observations it was established that the noise-level will vary at most 15\%  due to different background sources, which reduces the sensitivity for cosmic ray observations. However, due to their brightness these sources are not a common target for HBA observations.

\begin{figure*}%
\centering
\subfigure[][]{%
\label{fig:hba-beam-a}%
\includegraphics[width=0.45\textwidth]{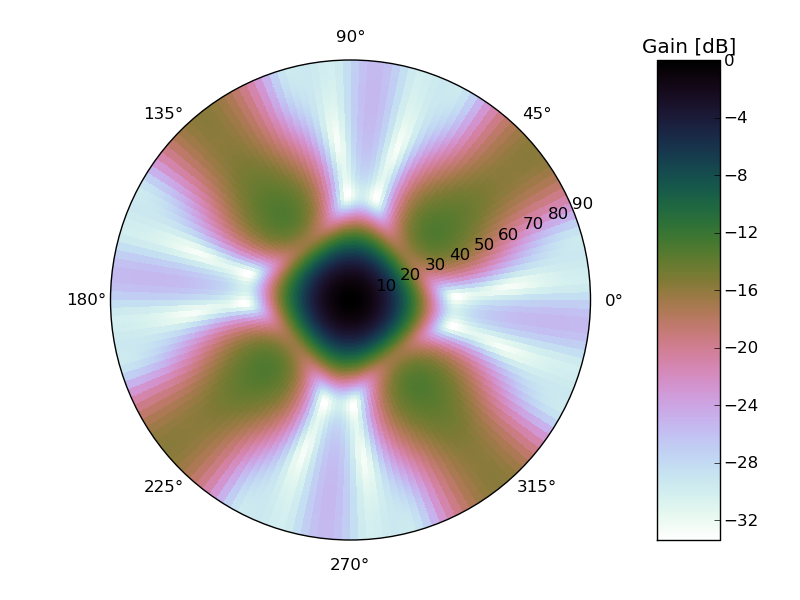}}%
\hspace{5pt}%
\subfigure[][]{%
\label{fig:hba-beam-b}%
\includegraphics[width=0.45\textwidth]{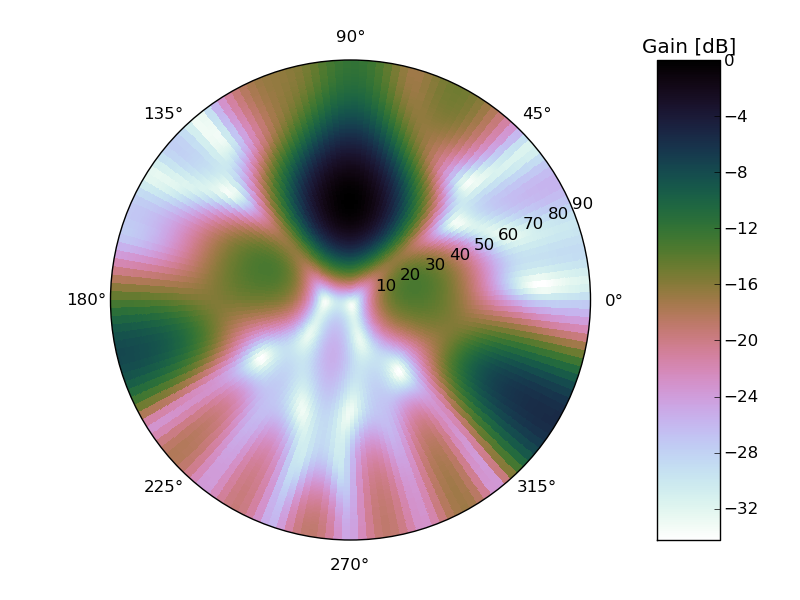}}\\
\subfigure[][]{%
\label{fig:hba-beam-c}%
\includegraphics[width=0.45\textwidth]{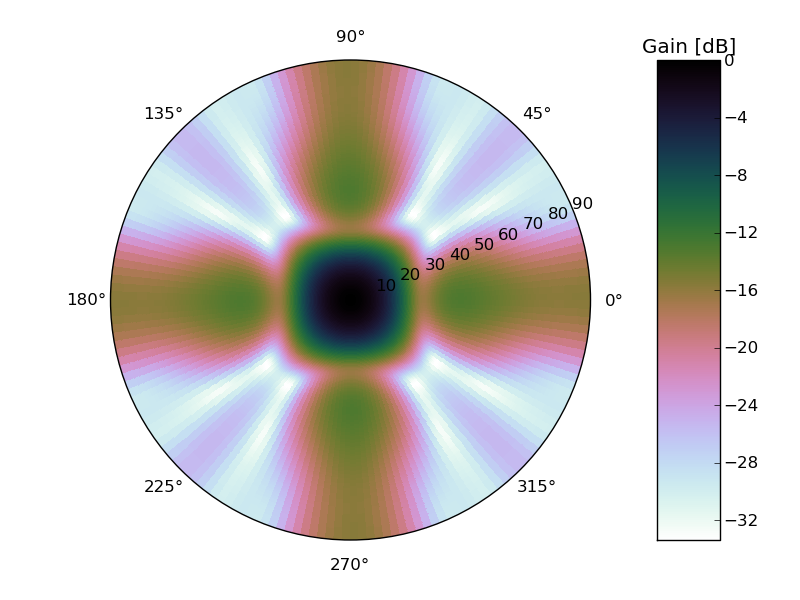}}%
\hspace{5pt}%
\subfigure[][]{%
\label{fig:hba-beam-d}%
\includegraphics[width=0.45\textwidth]{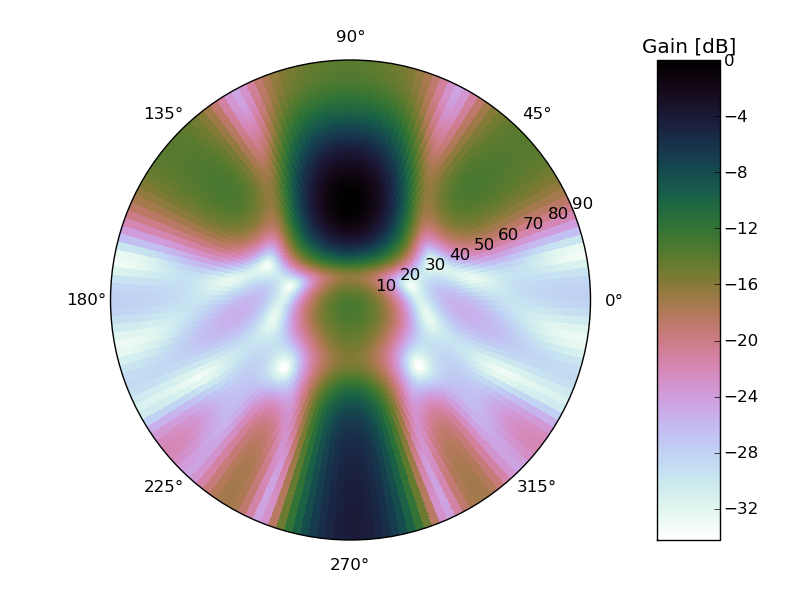}}\\
\subfigure[][]{%
\label{fig:hba-beam-e}%
\includegraphics[width=0.45\textwidth]{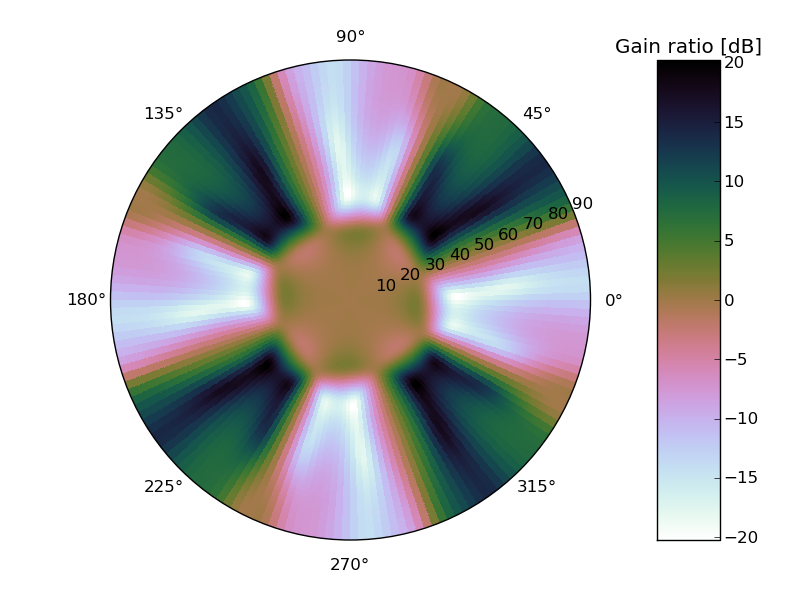}}%
\hspace{5pt}%
\subfigure[][]{%
\label{fig:hba-beam-f}%
\includegraphics[width=0.45\textwidth]{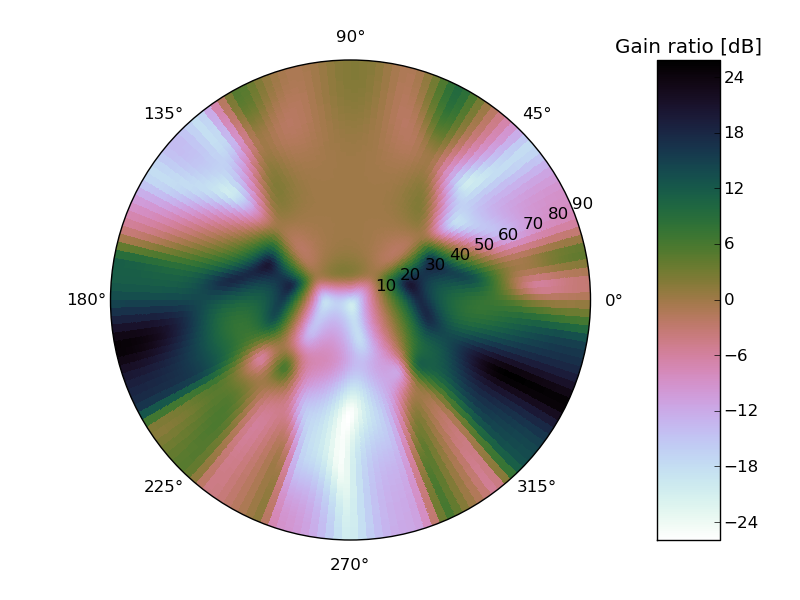}}\\
\caption{Influence of the beamforming applied at the HBA tile-level. Shown is the resulting gain in power as a function of cosmic ray arrival direction for two tile-orientations. They are given for a beam pointing towards local zenith in figures \subref{fig:hba-beam-a}, \subref{fig:hba-beam-c} and towards the North Celestial Pole in figures \subref{fig:hba-beam-b}, \subref{fig:hba-beam-d}. Figures \subref{fig:hba-beam-e} and \subref{fig:hba-beam-f} show the respective differences in gain between the two tiles from different sub-stations. These differences result in offsets between the measured signal strength in different sub-stations depending on the arrival direction of the cosmic ray.}%
\label{fig:hba_beams}%
\end{figure*}

\subsection{Gain corrections}
\label{sec:gain}
The HBA antennas are not read out individually but rather in tile-groups of 16 antennas \emph{after} analogue beamforming in the direction of the observation. In order to minimize artefacts in interferometric images, all individual HBA sub-stations are rotated at different unique angles (see section \ref{sec:instrument}). While the antenna elements within a tile are counter rotated by the same amount in order to observe the same polarization component, the grid of 16 elements is oriented with the sub-station orientation. This gives a different tile sensitivity pattern for each HBA sub-station, resulting in different gains between sub-stations within one single cosmic-ray event. Additionally, the non flat phase response due to the tile beamforming adds a direction dependent distortion to the pulse shape which is worse for cosmic ray arrival directions away from the observation direction.
To illustrate the complexities involved, figure \ref{fig:hba_beams} shows the influence of the analog tile-beamforming on a pure cosmic ray signal, without background noise, obtained from a CoREAS simulation \cite{CoREAS}. Here, the power gain
\begin{equation}
G_{P}=10\cdot\log_{10}\left(\frac{P_{\mathrm{out}}}{P_{\mathrm{in}}}\right)
\end{equation}
is given as a function of the cosmic-ray arrival direction for two HBA sub-station orientations and two beam directions. The simplest pattern is obtained for a beam pointing towards zenith, where the delay corrections are zero and the signals from the individual antennas are simply added (left-hand side in figure \ref{fig:hba_beams}). The gain pattern is in this case solely the result of delays introduced by the arrival direction of the cosmic ray. However, for many LOFAR observations the beam is not pointing towards zenith, but rather towards some astronomical object. A frequently occurring pointing is towards the North Celestial Pole, which is given as a second example (right-hand side). While the beam shapes for tiles in two sub-stations look similar, there are significant differences as depicted in the bottom row of figure \ref{fig:hba_beams}. These differences translate into differences in observed pulse amplitude of up to a factor of $\sim 15$ between tiles in two different sub-stations depending on the shower arrival direction. This means that while the beamforming always introduces an additional gain for cosmic rays arriving from the direction of the beam, the effect of the beamforming for off-beam cosmic rays will not be the same for every tile. While some signals might still be enhanced, others will be reduced to essentially noise-level. 

The exact differences depend on the shape of the pulse and on the frequency response of the electronics. Furthermore, the complex direction and frequency dependent response pattern of the individual elements needs to be taken into account as well. Crosstalk between antenna elements, due to the close spacing within a tile, requires a response pattern per element as the patterns will be slightly different. Such a detailed antenna model currently does not exist. Therefore, differences between tiles due to beamforming and the antenna pattern are not corrected for in the present analysis. 

In addition to the beam effect, there are intrinsic differences between stations and tiles. Gain differences between tiles within a station are corrected for using standard LOFAR calibration tables. These tables are generated regularly using the algorithms described in \cite{Wijnholds:2009,Wijnholds:2010}. The effect of possible gain differences between stations was tested, using data from the HBA part of the LOFAR MSSS survey \cite{MSSS}. The calibration values obtained form the pre-processed data of this survey vary between observations, but differ on average about 5\% between stations in any given observation. As these values are not stable on longer time-scales, they are not used to correct for gain differences in this analysis. This introduces a 5\% uncertainty on the pulse amplitudes measured in different stations.

Given the above mentioned lacking knowledge of the precise characteristics of the system, there is currently no absolute calibration for the electric field strength of the measurements.

\section{Dataset}
\label{sec:dataset}
Cosmic-ray data have been gathered with the HBAs since October 2011. Until November 2013, 155 events have been detected in the band of $\unit[110-190]{MHz}$. In addition, two events were detected in the band of $\unit[170-230]{MHz}$. The time spent observing the lower of the two HBA bands was about 20 times longer. Therefore, this article concentrates on the events measured in the lower band. 

The triggers from the scintillator array were sent according to the same specifications as for LBA observations \cite{Schellart2013}.  These settings give a threshold energy of $\unit[2\cdot 10^{15}]{eV}$ for the particle detection. While being recorded with the same trigger settings, the detection probability for an air shower based on its radio signal is found to be roughly a factor two lower for HBA than for LBA. This difference can be attributed either to an intrinsically reduced emission strength at higher frequencies or instrumental effects such as higher background levels and hardware differences.

\subsection{Information from particle data}
Every triggered radio event is complemented with parameters reconstructed from the particle data. For every event two reconstructed directions are available, one from the particle data and one based on the radio signals. The angular resolution of the particle detectors is on average ${1^\circ}$. Further parameters obtainable from the particle data are the position of the shower axis and an energy estimate of the primary cosmic ray. Both parameters are only reliably reconstructed for a certain parameter space \cite{Schellart2013}, and therefore not available for all events. The high quality events which are detected with the HBAs span an energy range from $\unit[1.7\cdot10^{16}]{eV}$ to $\unit[1.1\cdot10^{18}]{eV}$.
\subsection{Arrival directions}
\begin{figure}
\includegraphics[width=0.48\textwidth]{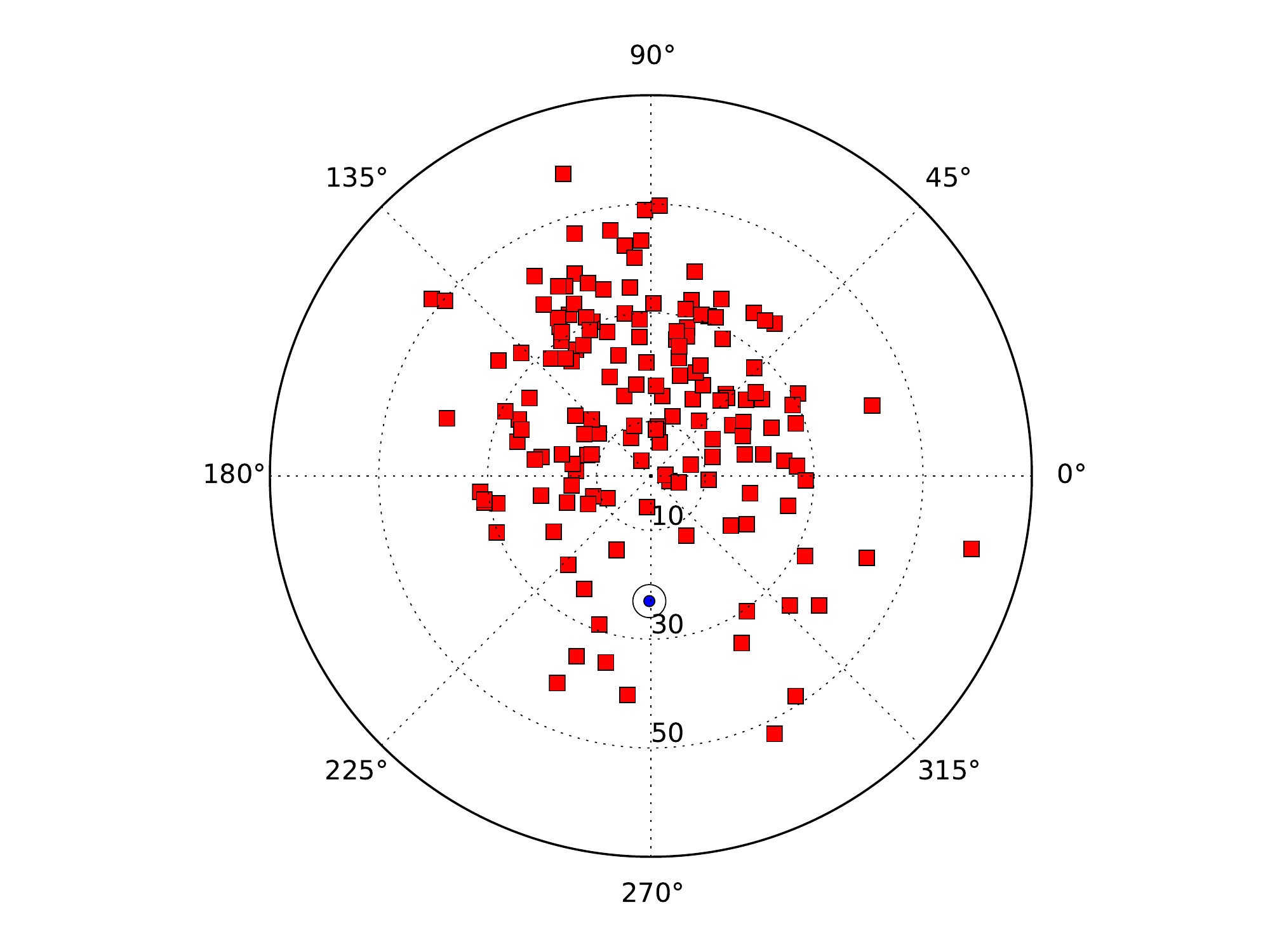}
	\caption[Skyplot]{Directions of the detected cosmic rays on sky as reconstructed from the particle data. $0^{\circ}$ corresponds to west and $90^{\circ}$ is north. The zenith angle ranges from $0^{\circ}$ to $70^{\circ}$. Also indicated (blue circle) is the direction of the magnetic field at the LOFAR core, which is pointing downwards to north. A clear asymmetry of number of detected events is visible.}
	\label{fig:skyplot}
\end{figure}
The arrival directions of the cosmic rays detected with the radio antennas are shown in figure \ref{fig:skyplot}. A clear north-south asymmetry is visible, which will be discussed in detail in section \ref{sec:NS}.

The angular resolution achieved with the HBA antennas is not the same for all directions. Many events are only measured with one station. As the antennas are clustered in two sub-stations, this results in a poorer angular resolution for showers arriving perpendicular to the axis connecting the two sub-stations. Also, the tile-beamforming has a negative effect on the accuracy  as it affects the pulse shape and thereby influences the reconstruction of the arrival time. Thus, in a similar analysis as presented in \cite{Schellart2013}, the angular resolution was determined to be $7^{\circ}$ with respect to the particle data. This angular resolution strongly decreases as a function of number of stations with detected pulses, but is on average worse than with the LBAs. Thus, a cosmic-ray candidate event is now accepted as a cosmic-ray event when the directions reconstructed from particle data and radio agree within $20^{\circ}$, instead of $10^{\circ}$. This relaxed cut excludes two obvious RFI candidates that arrived from close to the horizon and can also be identified by their deviating pulse form. This loosened cut provides larger statistics at a possible cost of lesser purity.

\subsection{Effect of the tile-beamforming}
Due to the statistical nature of the cosmic ray arrival directions, no event arrived directly ($< 1^{\circ}$) from the direction in which the beam of the observation was pointing. 

\begin{figure}
\includegraphics[width=0.48\textwidth]{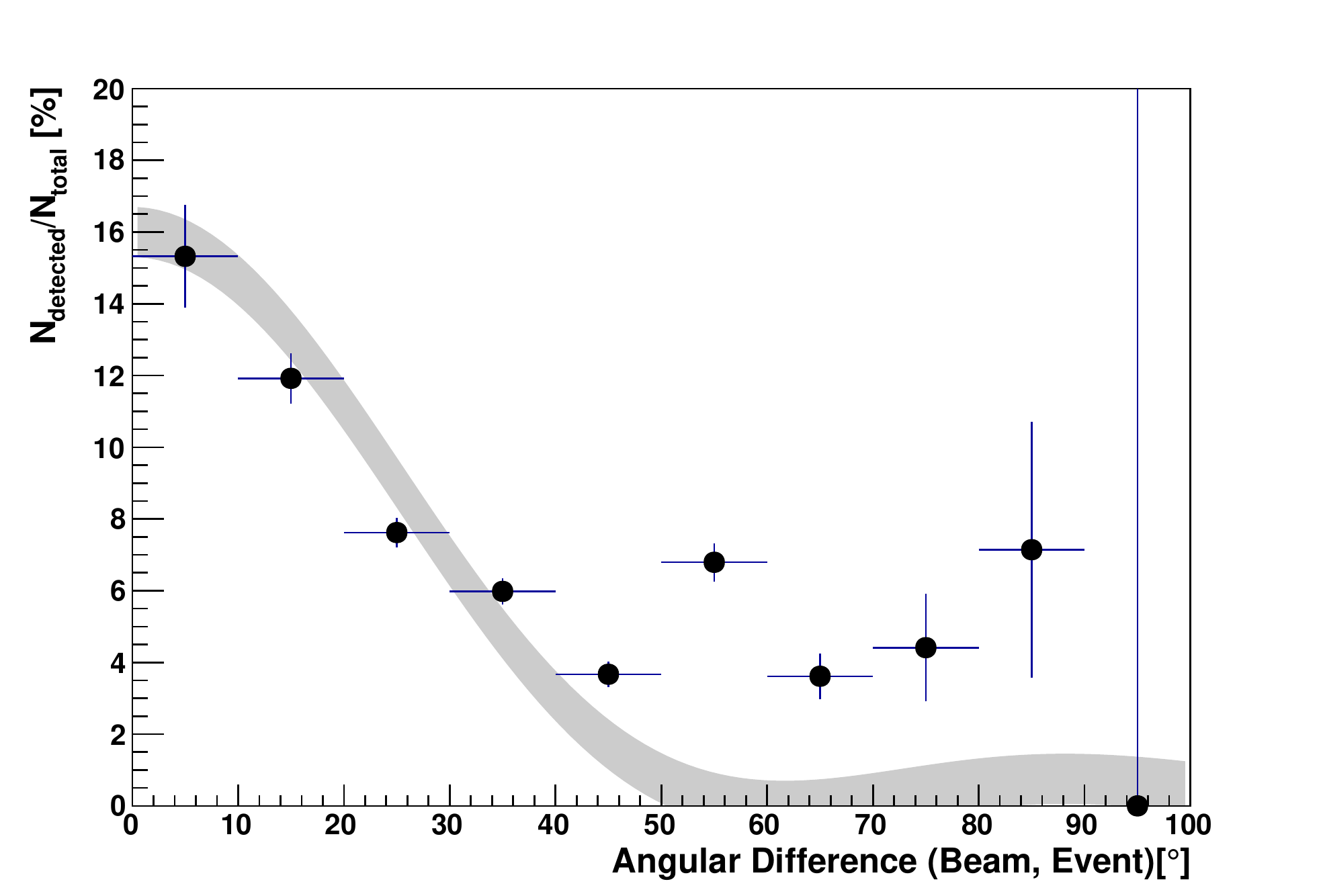}
	\caption[beamforming]{Probability of detecting an air shower as a function of angular difference between the direction of the observation and the arrival direction of the incoming cosmic ray. The detection probability is given per bin for all air showers with an arrival direction in this bin as a fraction of detected events, $N_{\mathrm{detected}}$, and received triggers $N_{\mathrm{total}}$. The grey contour shows a model of the extent of the beam-shape of an HBA tile. The model is scaled to match the cosmic ray data.}
	\label{fig:prebeam}
\end{figure}

The main effect of tile-beamforming in the direction of the shower is an increase in signal strength, which lowers the detection threshold. The main effect of beamforming in another direction than the arrival direction is a distortion of the pulse shape. This makes events of low signal strength harder to detect. Strong pulses are detectable, but the frequency content of the pulse as well as the position of the maximum will be affected. This effect is observed in data and visualized in figure \ref{fig:prebeam}, where the likelihood of detection is plotted as a function of angular distance between arrival direction and beam. The overall detection efficiency is determined by the energy threshold of the HBA tiles, which is higher than the threshold of the particle array However, the figure shows that events arriving closer to the beam direction are more likely to be detected and also that the detection probability does not go to zero with increasing distance. Interesting to note is that the distribution roughly follows the predicted dimensions of the beam of the HBAs. Using the relation for the diffraction pattern of an interferometer
\begin{equation}
G_\mathrm{Beam} \sim\frac{1}{\frac{\lambda}{D} \cdot \alpha} \cdot \sin\left(\frac{\lambda}{D} \cdot \alpha\right),
\end{equation} 
with typical wavelength $\lambda$ and detector size $D$, gives a full width half maximum beamwidth of about $\alpha = 20^{\circ}$ for an HBA tile and the distribution shown in figure \ref{fig:prebeam}. This beamwidth also describes the roughly $20^{\circ}$ region depicted figure \ref{fig:hba_beams} (e), in which the gain is independent of the rotation of the sub-station. Both beam effects essentially limit the field of view and sensitivity for cosmic ray observations with the HBAs.

\subsection{Observation of north-south asymmetry}
\label{sec:NS}
If the main contribution to the radio emission from cosmic ray air showers is geomagnetic in origin, a north-south asymmetry in the arrival direction of air showers measured in radio is expected \cite{Kahn:1966,Falcke:2003}. This has indeed been observed by many experiments in the frequency range up to $\sim\unit[100]{MHz}$ \cite{Falcke:2005,Codalema2009}. If this still holds true for the frequency range considered in this paper ($\unit[110-190]{MHz}$), such a north-south asymmetry should also be visible in Fig.~\ref{fig:skyplot}. This indeed is the case. However, for the particular setup at LOFAR there is an additional complication. As the sensitivity of the instrument depends on the arrival direction of the cosmic ray relative to the current pointing of the tile-beam, an observed asymmetry in air shower arrival directions might be the result from an asymmetry in the beam pointing rather than caused by the intrinsic air shower radio emission process.

\begin{figure}
\includegraphics[width=0.48\textwidth]{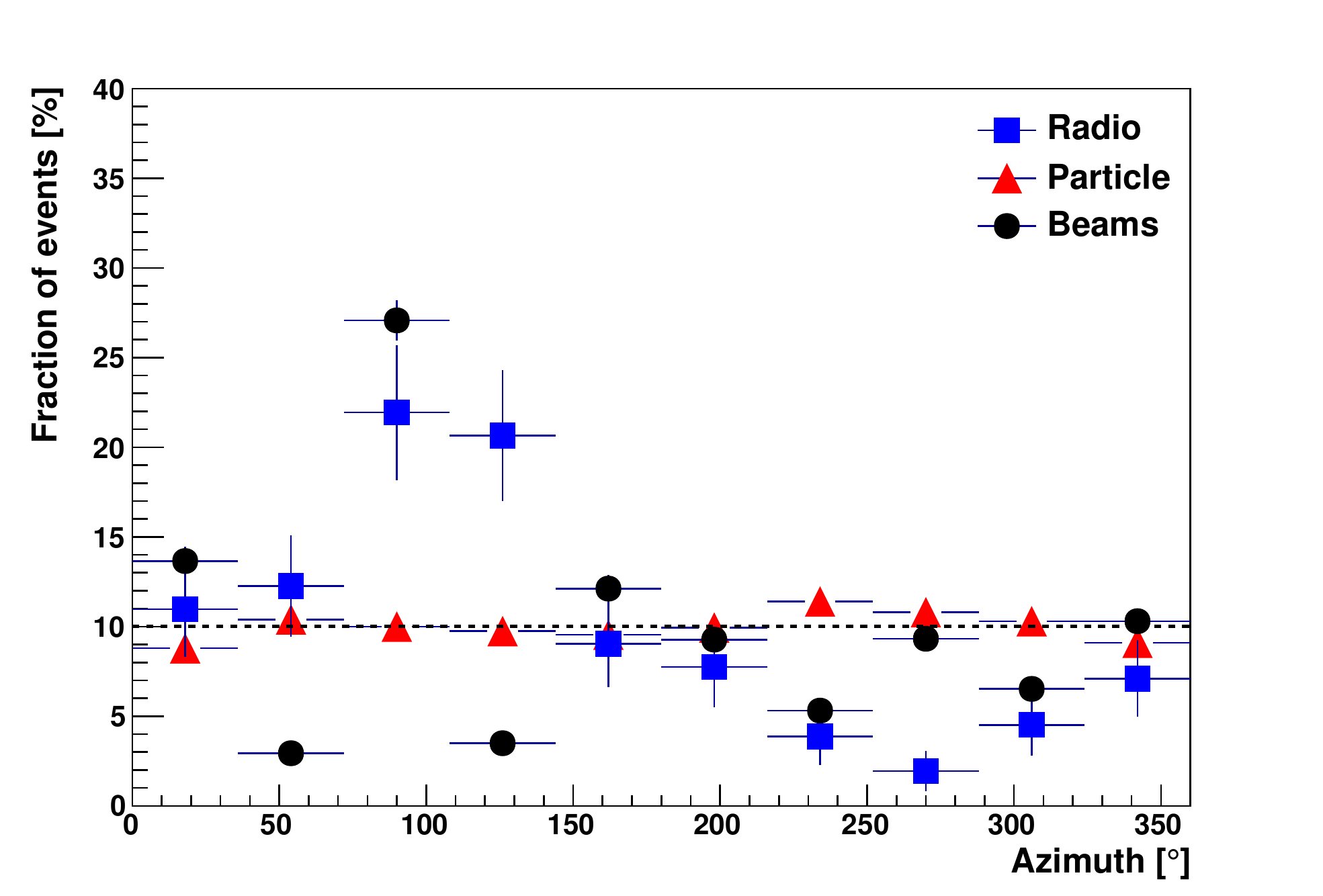}
	\caption[Angles]{Distribution of arrival directions of the measured air showers with respect to azimuth angle. The azimuth angle is measured northwards positive from east. Shown are the different distributions of azimuth angles of the direction in which the data were tile-beamformed (black circles), the directions of the air showers that triggered LOFAR (red triangles) and the directions of the cosmic rays, which were detected in the radio (blue squares). The triggers are almost uniformly distributed while the beams of the astronomical observations and the radio detections are not.}
	\label{fig:angles}
\end{figure}

In Fig.~\ref{fig:angles} it can be seen that while the reconstructed arrival directions of the cosmic rays detected by the particle detectors are uniformly distributed in azimuth, the subset of those triggers that had a detectable radio signal are not. In the same figure the tile-beam directions for all triggered events are also indicated. Although at first glance, the distribution of radio events seems to follow the beam direction distribution, a closer inspection does show some important differences. In the second bin for example, the fraction of detected radio events is much larger than the fraction of beams pointing in this direction. It is important to stress here again that such a discrepancy is possible since while the sensitivity is higher in the beam direction, it is not zero outside of the beam. Thus, a cosmic ray coming from outside the beam can still be detected if the signal is sufficiently strong.

A total of 155 cosmic rays were detected. Of those cosmic rays 116 arrived from the northern half of the hemisphere and 39 from the southern half giving a ratio $N/S = 3.0$.  In order to check if this asymmetry can be explained solely by the asymmetric distributions of tile-beams (a ratio of 1.46 for all beams, resulting in 92 north vs. 63 south for 155 observations), a simple Monte Carlo procedure is followed. For each trial the number of events arriving from north and south are drawn from a Poissonian distribution with expectation values 92 and 63 respectively. The ratio of events from north over events from south is calculated. This ratio follows a neither Poissonian nor Gaussian distribution. A total number of $10^7$ trials are performed to give the probability distribution of the north over south ratio. This procedure was repeated 100 times giving a total number of $10^9$ trials. The cumulative probability of observing a ratio of $3.0$ or higher is $P(r\geq 3.0) = 3.4 \cdot 10^{-5}$ on average. In order to reduce the probability to be dominated by events arriving close to the north-south boundary or near the zenith, where north and south are not defined, the same analysis is repeated on a set that excludes these regions by $\pm5^\circ$. This exclusion region is motivated by the beamwidth and excludes 7 air showers. The remaining air showers are distributed in 112 from north and 36 from south with a fraction of 1.57 from north for all beams remaining in the selected regions. The resulting probability for this set is found at a value of  $1.9  \cdot 10^{-4}$. 

Furthermore, if we only select the 61 beams that were actually pointing to the south, we still detect 38 air showers in them arriving from the north. The result also does not change if the arrival directions reconstructed from LOFAR data itself are used instead of the ones from particle detector data.

It is thus extremely unlikely that the observed asymmetry in cosmic ray arrival directions is caused by the asymmetry of the instrumental sensitivity alone. We therefore conclude that the distribution of arrival directions of air showers measured at $\unit[110-190]{MHz}$ is compatible with a strong geomagnetic component in the emission.

% -----------------------------------
% -----------------------------------

\section{Observation of Cherenkov rings in air showers}
\label{sec:discussion}
The high antenna density of LOFAR enables detailed studies of the radiation pattern generated by individual showers. This is very instructive due to the intrinsic asymmetry of the signal which hinders averaging over showers. LOFAR is the only current experiment that can test theoretical predictions about the signal pattern in a single event study.

Using these single events several approaches how to extract information from the radio pattern about the characteristics of the air shower are tested. The measured patterns are in their full extent compared to Monte Carlo simulations. This illustrates to what extent the current air shower simulations capture the measured features and how this contains information about the shower development. Furthermore, the measured events show the predicted ring structures and these are used to extract the ring size. Here, we also discuss with what precision the ring size can be determined and whether this is sufficient for a measurement of the height of the shower maximum. 

\subsection{Example event}
\begin{figure}
\includegraphics[width=0.5\textwidth]{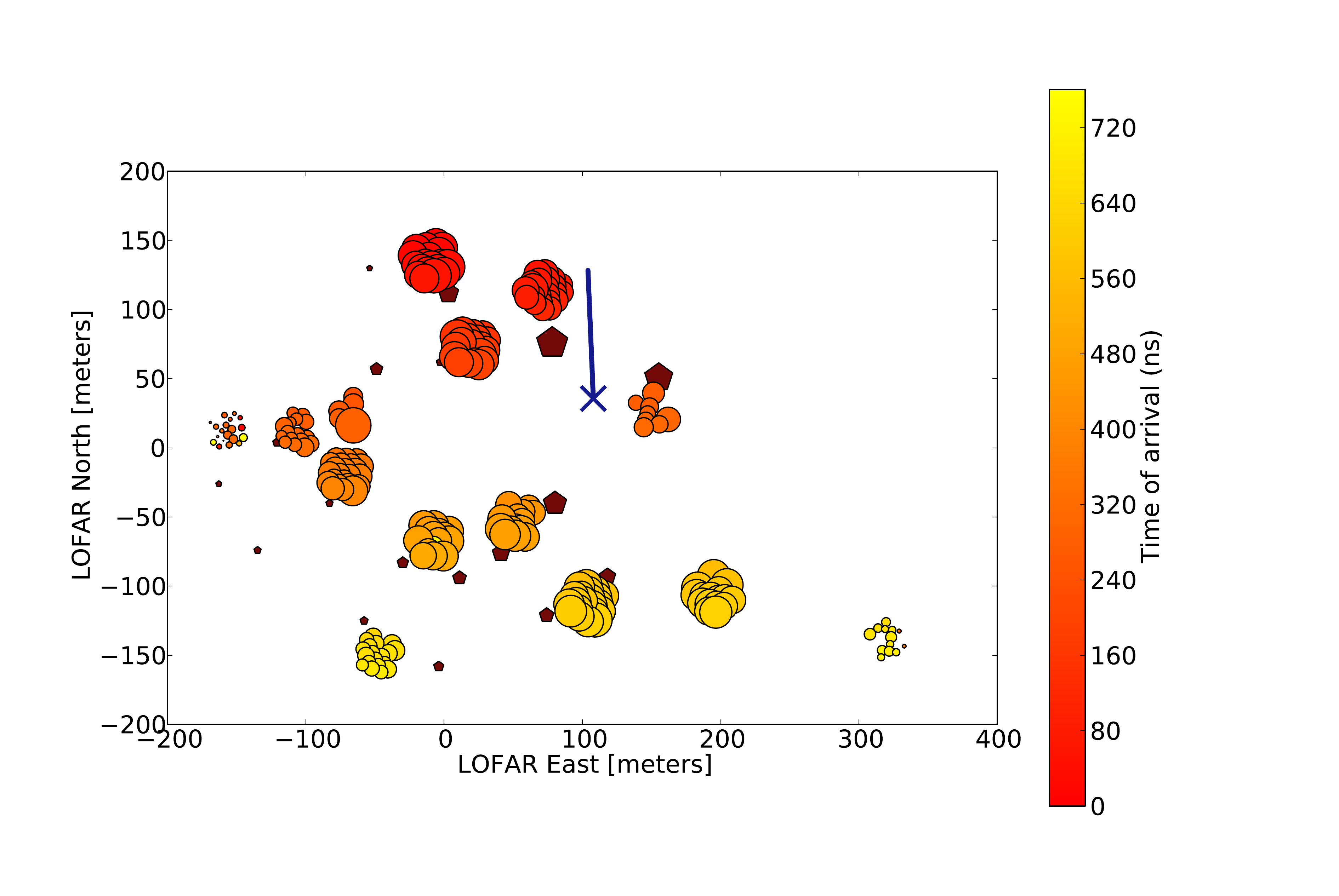}
	\caption[Footprint]{Air shower as measured with the HBAs. The dark red pentagrams depict the particle detectors. The colored circles represent the HBA tiles. In both cases, signal strength (particle density and maximum amplitude) is encoded in the size of the markers (logarithmic and linear, respectively). The arrival times of the radio pulses are encoded in the color of the circles going from early in red to late in yellow. The shower geometry as reconstructed from the particle data is shown with the blue cross and the line indicating the shower core and the projected arrival direction, respectively. This air shower is also used for further analysis in section \ref{sec:comp}.}
	\label{fig:footprint}
\end{figure}

Figure \ref{fig:footprint} shows an event as recorded with the HBAs. The distribution of signal strengths on the ground shows the level of detail in which the features of the emission can be measured with LOFAR. It is however difficult to identify general structures by eye, given the irregular distribution of antennas and the measurement of the radiation pattern on the ground instead of in the shower plane. 

\subsection{Comparison to simulations}
\label{sec:comp}
\begin{figure}
\includegraphics[width=0.5\textwidth]{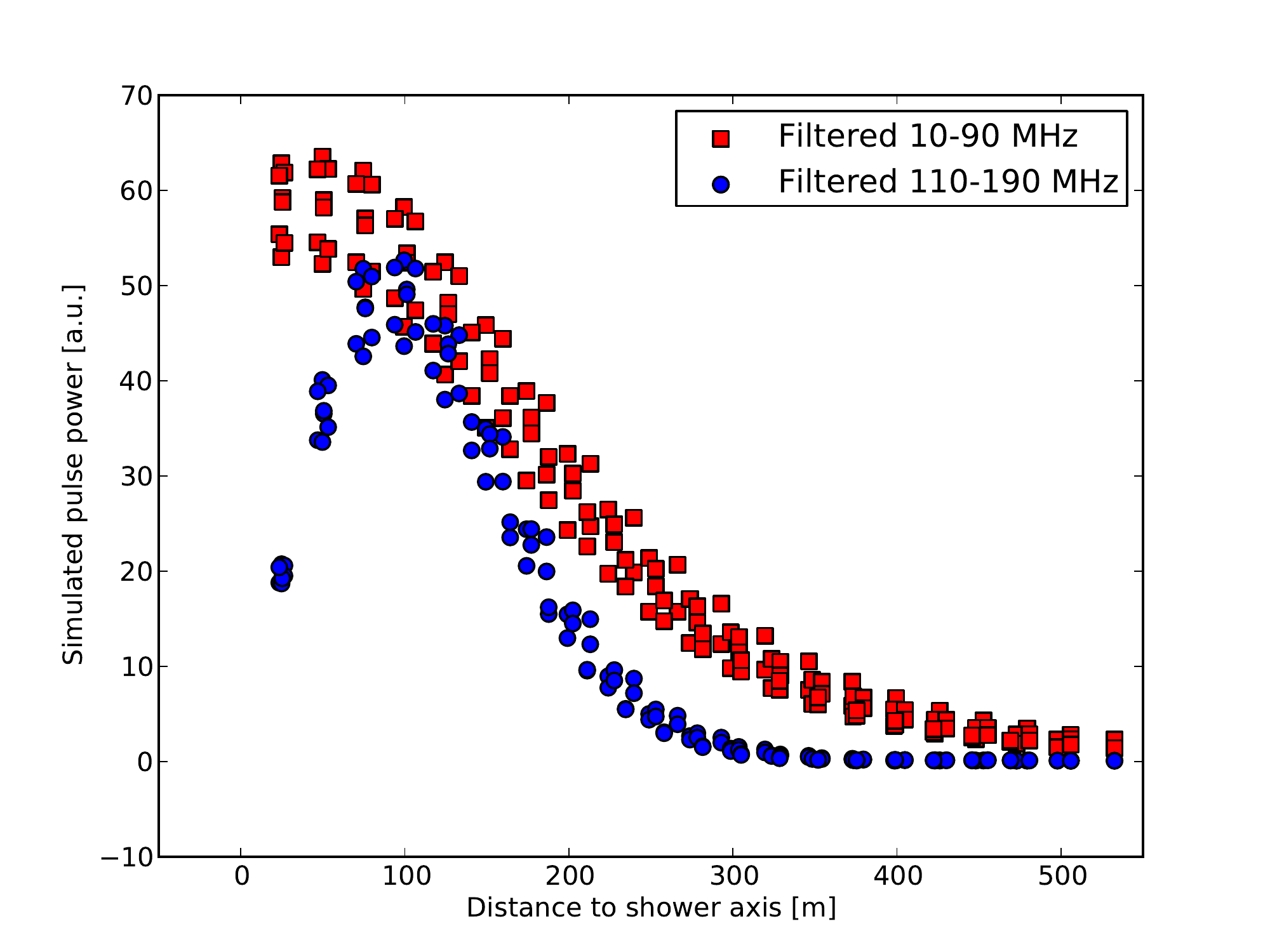}
	\caption[Filter]{Simulated pulse power (CoREAS) as a function of distance to shower axis for different ideal filter settings on an idealized grid of antennas. For this shower the primary particle was a proton of $7.2\cdot 10^{17} $eV, arriving at a zenith angle of $38^{\circ}$. As the pulse power is not only a function of distance to shower axis, the spread represents the asymmetry in the signal. }
	\label{fig:filters}
\end{figure}

As shown in \cite{Buitink2014} the simulation code CoREAS \cite{CoREAS} describes the data from the low-band antennas well and in great detail. With these simulations it can be illustrated what is expected from an observation at higher frequencies. Figure \ref{fig:filters} shows the same simulation filtered in the two different LOFAR frequency bands. The shape of the signal distribution as function of the distance to the shower axis changes at higher frequencies. A ring like structure emerges, which corresponds to the enhancement at around $\unit[100]{m}$. Furthermore, the total power of the signal decreases with frequency.

A direct comparison of the HBA data using the method of \cite{Buitink2014} can only be accomplished under certain conditions. As discussed earlier, the additional gain differences of the HBA sub-stations according to their rotation make it challenging to correct for the hardware response. To do this correctly, one has to simulate single pulses and feed them through a full model of the hardware, including especially the analogue beam former. Such a realistic model of the full hardware is however not available yet.

Instead, we can concentrate on events that arrived from close to the direction of the beam. As shown in figures \ref{fig:hba_beams} and \ref{fig:prebeam} the response of all HBA sub-stations is similar for events arriving from the full-width-half-maximum (FWHM) of the main beam. We selected three such events with more than four triggered stations and compared them to simulations. All three events arrived from close to the north celestial pole.

For the comparison we employ the same method as described in \cite{Buitink2014}. Each event is characterized by its arrival direction (reconstructed from the arrival times of the radio pulses) and energy (reconstructed from the particle detections). The CoREAS~\cite{CoREAS} plugin of CORSIKA~\cite{Heck:1998}, employing QGSJETII.04~\cite{Ostapchenko:2006} as the hadronic interaction model, is used to generate radio emission intensity patterns for the given input parameters. Due to shower-to-shower fluctuations the generated pattern will be different for each simulated shower. The main factor determining pattern differences is $\mathrm{X}_{\mathrm{max}}$, the distribution of which depends on the type of the primary particle. Thus, 25 showers are generated with a proton as the primary and 15 with an iron nucleus using the energy estimate and arrival direction from the particle data as input. The predicted radio intensity pattern for each shower is compared with the measured data. Free parameters in this least-squares fit are the position of the shower axis and a single scaling factor, compensating for the missing absolute calibration. The simulation with the best matching pattern is selected as the comparison event.

\begin{figure*}%
\centering
\subfigure{%
\includegraphics[width=0.43\textwidth]{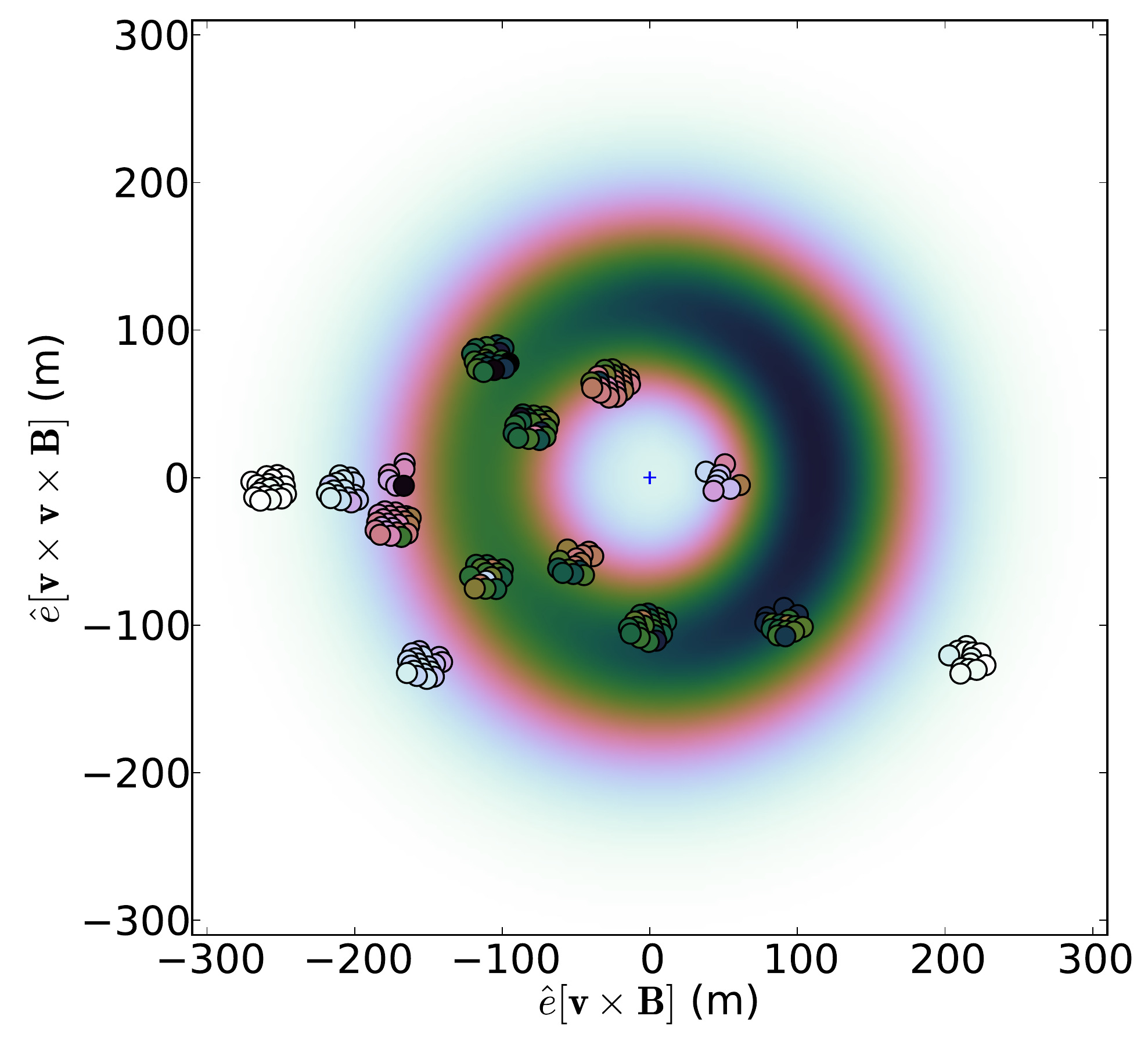}}%
\hspace{5pt}%
\subfigure{%
\includegraphics[width=0.45\textwidth]{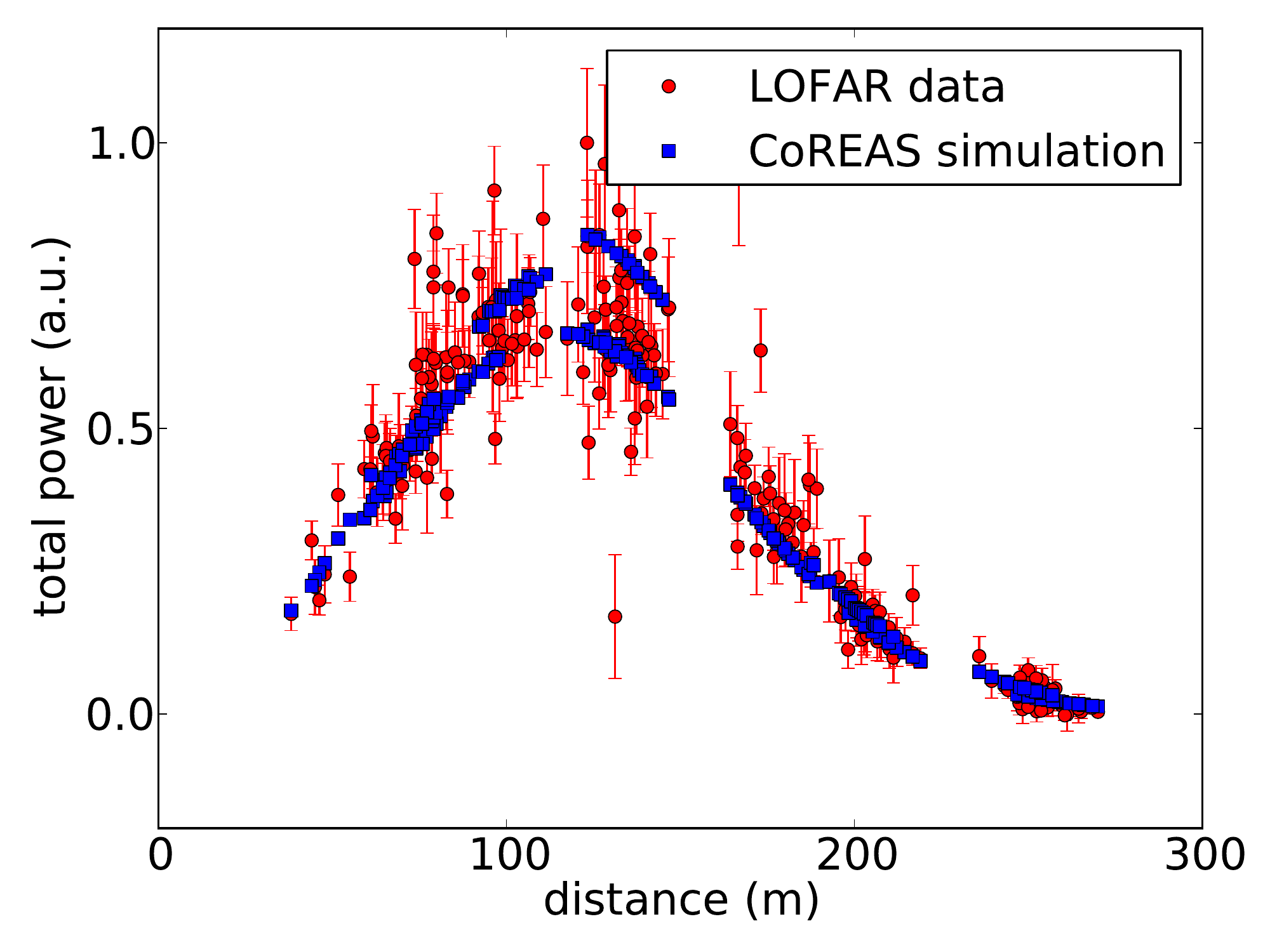}}
\subfigure{%
\includegraphics[width=0.43\textwidth]{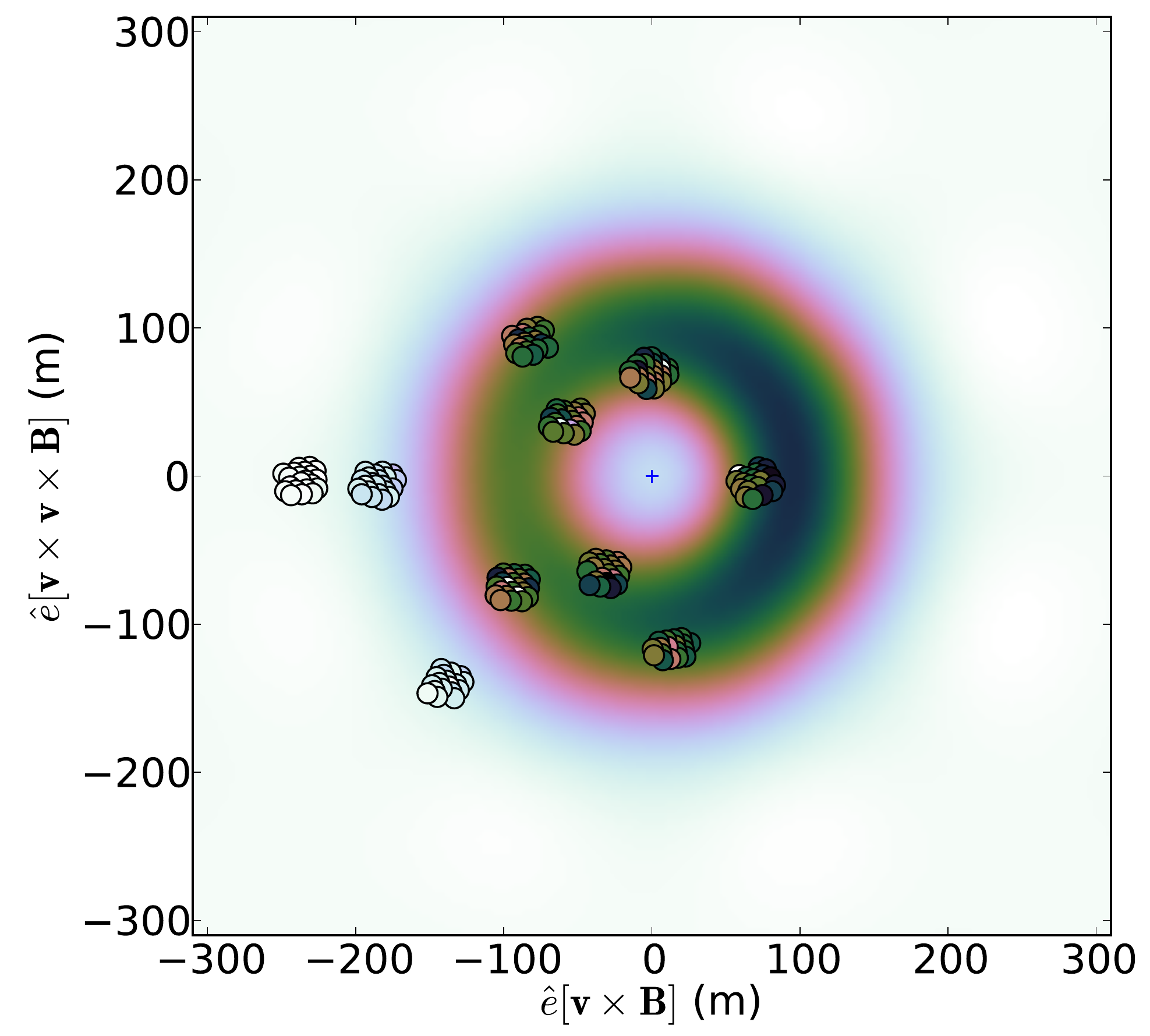}}%
\hspace{5pt}%
\subfigure{%
\includegraphics[width=0.45\textwidth]{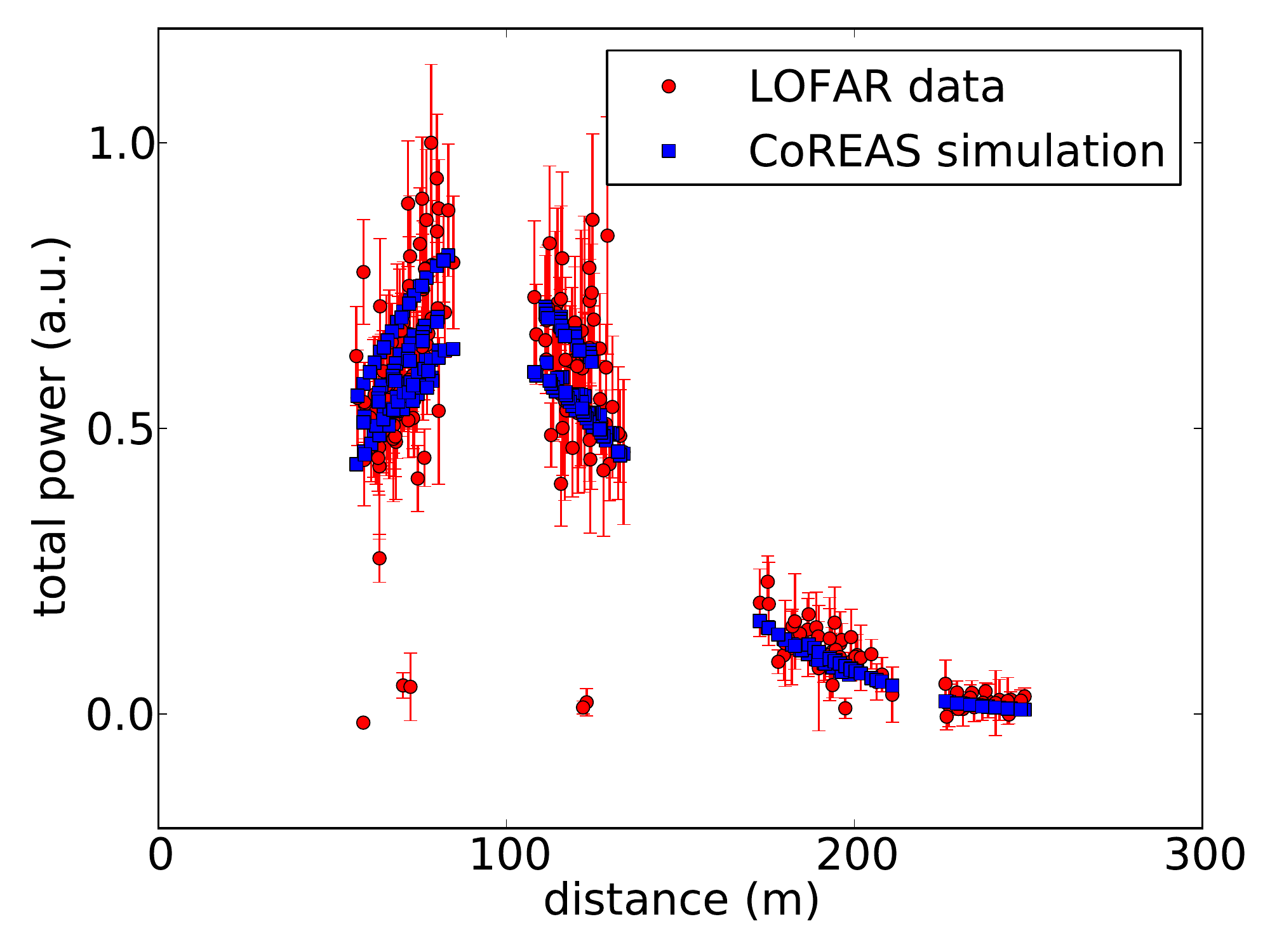}}
\subfigure{%
\includegraphics[width=0.43\textwidth]{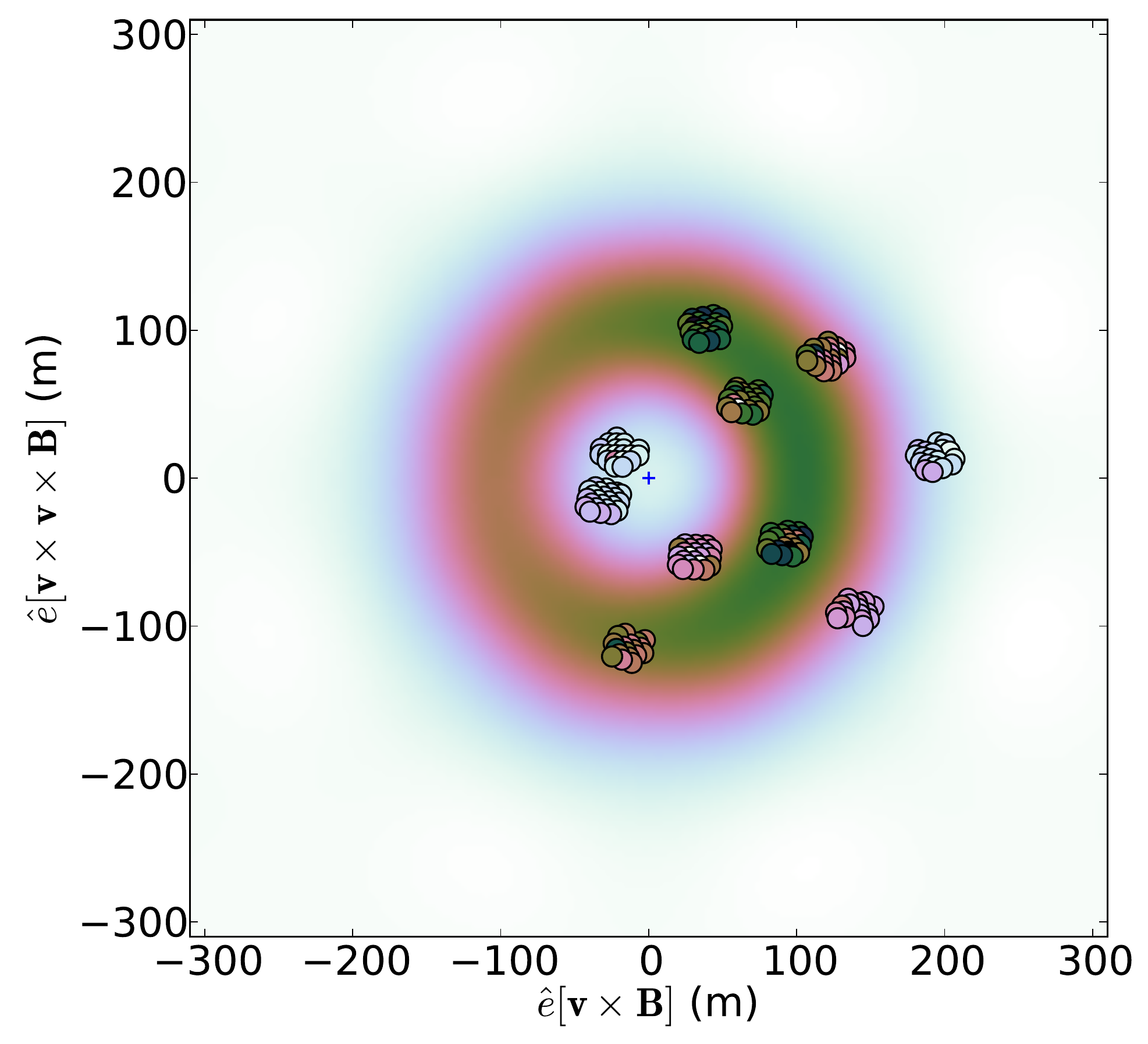}}%
\hspace{5pt}
\subfigure{
\includegraphics[width=0.45\textwidth]{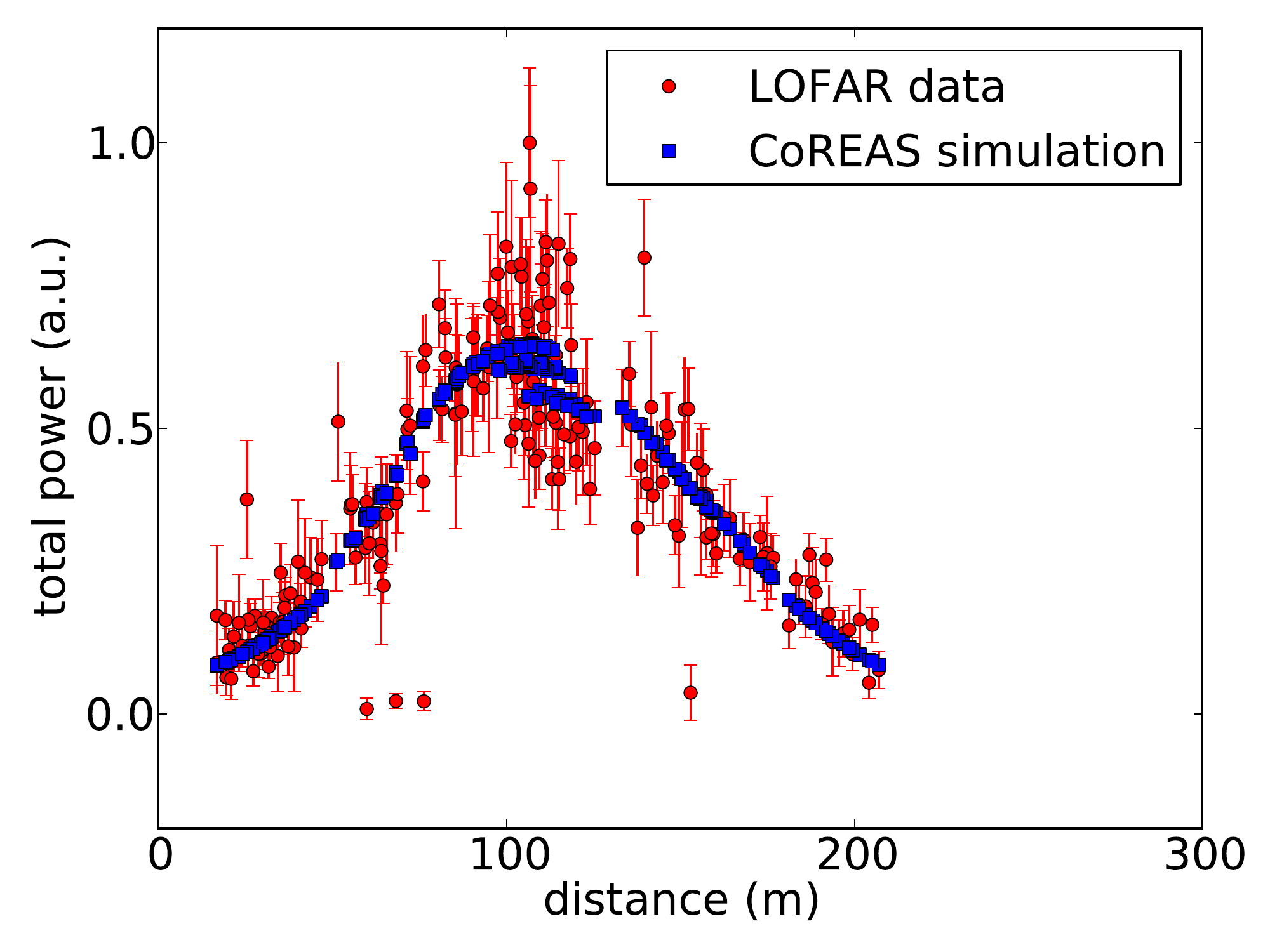}}
\caption{Comparison of measurements and simulations for three air showers. Left: Signal distribution in shower plane. The circles indicate the positions of all measured signals from tiles operational at that moment. The background map is the interpolated signal strength from the best fitting simulated CoREAS shower. The integrated power from $\unit[110-190]{MHz}$ both for measurements and simulations is encoded in color. The reference coordinate system is the shower plane defined by the propagation velocity vector of the shower \textbf{v} and magnetic field direction \textbf{B}. The shower axis is located at the plus sign. Right: Corresponding integrated radio pulse power for simulation (blue squares) and LOFAR HBA measurements (red circles) as a function of distance to shower axis.}
\label{fig:simulation}
\end{figure*}

The best fitting simulations for the three events are shown in figure \ref{fig:simulation}. In the images on the left-hand side, the interpolated total power from the simulation is given in the background map. Overlaid are the measured data as circles. Where the colors match there is an agreement in signal strength. The Cherenkov ring dominates the structure for all events, which is clearly visible in both simulations and measurements.

The same can be seen in a projection of the signals as a function of distance to shower axis as shown on the right-hand side of figure \ref{fig:simulation}. This lateral distribution is dominated by the amplified ring structure at distances of about 100 meters. These measurements confirm the importance of the propagation of the radiation, which causes these relativistic time compression of the measured emission, which itself is still dominated by the geomagnetic effect (see \ref{sec:NS}).

\subsection{Sensitivity of the Cherenkov ring to the depth of the shower maximum}
\label{sec:sens}
It was suggested in  \cite{de-Vries:2013} that the radius of the Cherenkov ring is sensitive to the depth of shower maximum $X_{\mathrm{max}}$.  Experimentally, it has to be tested how precisely the radius of the ring can be measured and what resolution this yields for $X_{\mathrm{max}}$.

The three events introduced above are used to determine the precision with which the radius of the Cherenkov ring can be obtained. The ring size is defined as that distance to the shower axis at which the pulse power reaches its maximum value. This distance of the highest signal strength is obtained by fitting a Gaussian function to the distribution of pulse power as function of distance to the shower axis. 

When applying this method a number of uncertainties have to be taken into account. The main contribution to the overall uncertainty is the uncertainty on the position of the shower axis. Unless there is a complete fit of the signal distribution of the radio data \cite{Nelles2014}, the axis as obtained with the particle array has to be used. The array delivers uncertainties on the axis position that vary from 5 meters to 30 meters and on the arrival direction of about $1^{\circ}$.

In order to account for these uncertainties, the signal strength as a function of distance to the shower axis is recalculated 500 times for a shower axis that is varied within its uncertainties. The resulting ring sizes obtained by fitting the Gaussian to the new distributions are filled into a histogram. From this histogram the most probable value of the ring size and the corresponding uncertainty can be obtained. 

As can be seen in figure \ref{fig:simulation}, the signal pattern is not completely axis-symmetric. This is not taken into account, when using a one-dimensional fit to describe the whole distribution. It seems, however, that the radius of the ring is sufficiently symmetric in the shower plane to approximate this feature in one dimension, and thus a fit of a Gaussian function is chosen to determine the ring size.

The parameters reconstructed from the radio pulse powers of the three example showers are shown in table \ref{tab:par}.
\begin{table}
\begin{tabular}{lcc}
\hline
Event&Zenith angle [$^\circ$]&Radius of ring [m]\\
\hline
\hline
1&$43.4\pm2.0$&$117.3 \pm 4.7 $\\
2&$34.9\pm1.0$&$93.3 \pm 2.1$\\
3&$40.5\pm1.0$&$119.6 \pm 22.1$\\
\hline
\end{tabular}
\caption{Shower parameters determining the measurement of $X_{\mathrm{max}}$. The angle is measured from zenith (upwards: $0^{\circ}$) and reconstructed from radio data. The arrival directions are compatible with the ones reconstructed from particle data.  The ring size is determined according to the procedure described in section \ref{sec:sens}. }
\label{tab:par}
\end{table}

In \cite{de-Vries:2013} $X_{\mathrm{max}}$ is calculated by fitting the following relation to a set of air showers, simulated using the EVA code \cite{EVA}, with energies between $\unit[10^{17}-10^{19}]{eV}$:
\begin{equation}
 X_{\mathrm{max}} = a + b \cdot d_{\mathrm{max}}.
\label{eq:xmax}
\end{equation}
The parameters $a$ and $b$ are fitted constants and $d_{\mathrm{max}}$ denotes the radius of the ring. 

\begin{figure}
\includegraphics[width=0.5\textwidth]{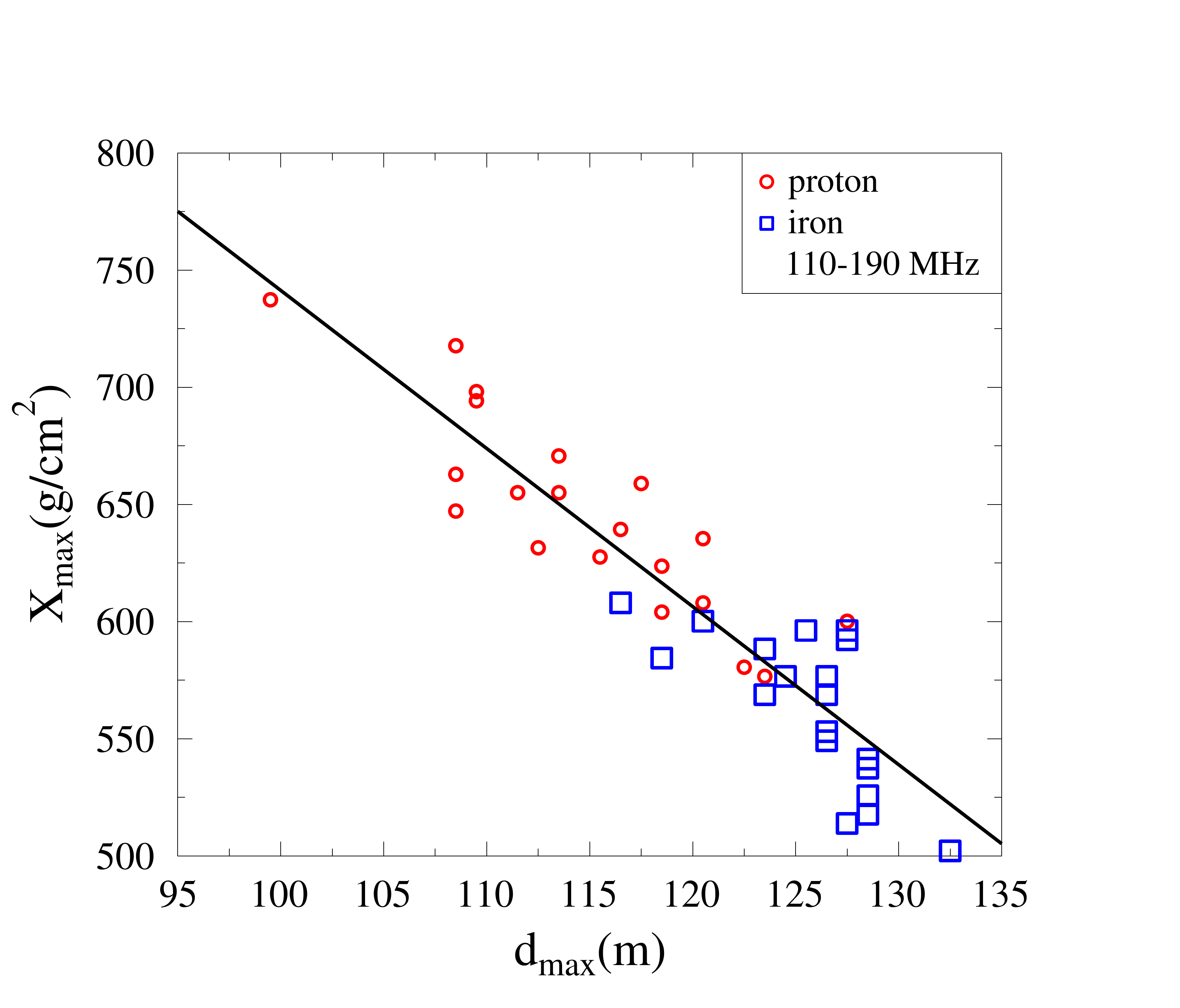}
	\caption[EVA]{Distribution of ring sizes obtained from EVA simulations for different values of the height of the shower maximum. Air showers were simulated with iron (blue squares) and proton primaries (red circles), a zenith angle of $45^{\circ}$ and a random azimuth angle. The radio signals are filtered to match the frequency range of the HBAs. The magnetic field configuration was the one of the center of LOFAR. The height of the shower maximum can be expressed by a linear function of the ring size, which is indicated by the fit. }
	\label{fig:eva_new}
\end{figure}

In \cite{de-Vries:2013} the generated air showers are vertical and the configuration of the magnetic field is different in both direction and strength from that at the central site of LOFAR. This makes it necessary to redo the calculations to obtain a prediction for the relation of the ring size to the height of the shower maximum appropriate for LOFAR observations. The same analysis for showers of a more typical arrival direction ($\theta=45^{\circ}$) in the frequency range of $\unit[110-190]{MHz}$ with the magnetic field configuration at the LOFAR location, delivers the results shown in figure \ref{fig:eva_new}.

For vertical showers and frequencies below $\unit[200]{MHz}$, \cite{de-Vries:2013} find a double-peaked structure in the distribution of signal strengths, which complicates the usage of the ring size as tracer of the shower maximum. For the magnetic field configuration of LOFAR and the higher inclination angle this feature does however not occur. Therefore a linear relation between ring size and shower maximum can still be used, as evidenced by Fig.~\ref{fig:eva_new}. 

For a perfectly measured radius of the ring a resolution of better than $\unit[25]{g/cm^2}$ is achieved. Combining this uncertainty with the uncertainties for the reconstruction of the ring size, yields the results as shown in table \ref{tab:xmax}. The results are compared to the best fit obtained from CoREAS simulations, as described above. 

\begin{table}
\begin{tabular}{lcc}
\hline
Event&Fit: $X_{\mathrm{max}}$ [g/cm$^2$]&Sim: $X_{\mathrm{max}}$ [g/cm$^2$]\\
\hline
\hline
1&$45^{\circ}:631\pm\ 85$&$675\pm$ 22\\ 
2&$30^{\circ}:548\pm\  58$&$643\pm$ 27\\ 
3&$45^{\circ}:609\pm156$&$671\pm$ 37\\

\hline
\end{tabular}
\caption{The height of the shower maximum $X_{\mathrm{max}}$ as obtained for the example events. The methods are using a parametrization for different zenith angles (Fit) or dedicated individual simulations (Sim).}
\label{tab:xmax}
\end{table}

This shows that there is, for the given three events, no statistically significant discrepancy between the value for $X_{\mathrm{max}}$ obtained with the two methods. However, the precision obtained by directly comparing to simulations is much higher than the one from using only the ring size. It should also be noted, that one cannot use the discrepancies between the values to compare the two models (EVA and CoREAS), as two completely different methods are applied to obtain $X_{\mathrm{max}}$.

There are a number of additional uncertainties that need to be considered when comparing the results. As shown in \cite{de-Vries:2013} the frequency range in which the shower is measured has a strong effect on the location of the ring. As there is no full antenna model to correct the HBA observations to a fully flat frequency spectrum, this might bias the results. Also, the missing antenna correction influences the $X_{\mathrm{max}}$ obtained with direct comparison to simulations. In contrast to measurements with the low-band antennas, the hardware response for the high-band antennas could not be included. It also should be noted that there is no parametrization of the ring size as a function of zenith angle yet and the events are approximated by the closest available set of simulations. 

We show that the radius of the Cherenkov ring can be used with experimental data as an indicator for the depth of the shower maximum. However, the obtainable accuracy is far less than needed for a precise composition study. It is therefore necessary to use more information than just the ring size. This information is available in the fact that the distribution of the signal is a non-symmetric function of several shower parameters. Employing a more complex fitting procedure \cite{Nelles2014} or a direct comparison to simulated showers \cite{Buitink2014} uses this information.

\section{Conclusions and Outlook}
The data of air showers collected with the LOFAR high-band antennas provide unprecedented detailed measurements of radio emission in the frequency range of $\unit[110-230]{MHz}$. In standard observation mode, 155 cosmic rays were measured between October 2011 and November 2013.

The high-band antennas were designed to observe (astronomical) objects in pre-defined directions. The effect of the analogue beamforming is difficult to remove completely for cosmic-ray measurements. It influences the absolute calibration between groups of high-band antennas and affects the shape of the pulses. This makes the high-band antennas a less optimal tool than the low-band antennas. However, we show that it is worthwhile to attempt a detailed correction of the hardware effects. 

For the first time, we measure a dominant relativistic time compression of the radio emission of air showers on a single-event basis in the frequency range of $\unit[110-190]{MHz}$. We show that it is possible to measure the radius of this Cherenkov ring with an accuracy of less than 20 meters. This is sufficient to give an indication of the depth of shower maximum. However, more complex procedures are needed to resolve the shower maximum with the necessary accuracy for composition studies. Given that LOFAR is observing with the high-band antennas roughly 50\% of the time, reconstructing the HBA data to a better quality will significantly increase the event statistics for composition studies at LOFAR.  

Also, measuring the same air shower with both types of antennas is very promising. The well-understood measurements with the low-band antennas could be extended to higher frequencies to learn more about the emission mechanisms. The current measurements strongly encourage to implement this mode of combined observation in the complete frequency range from $\unit[10-230]{MHz}$ in LOFAR.

\section*{Acknowledgements}
We thank both the internal and external reviewers for the fruitful discussions regarding this article. The LOFAR cosmic ray key science project very much acknowledges the scientific and technical support from ASTRON. Furthermore, we acknowledge financial support from the Netherlands Research School for Astronomy (NOVA), the Samenwerkingsverband Noord-Nederland (SNN), the Foundation for Fundamental Research on Matter (FOM) and the Netherlands Organization for Scientific Research (NWO), VENI grant 639-041-130. We acknowledge funding from an Advanced Grant of the European Research Council under the European Union's Seventh Framework Program (FP/2007-2013) / ERC Grant Agreement n. 227610.

 LOFAR, the Low Frequency Array designed and constructed by ASTRON, has facilities in several countries, that are owned by various parties (each with their own funding sources), and that are collectively operated by the International LOFAR Telescope (ILT) foundation under a joint scientific policy.
\bibliographystyle{elsarticle-num}
\bibliography{BIB_LOFAR}

\end{document}